\documentclass[%
 reprint,
 amsmath,amssymb,
 aps,
]{revtex4-2}

\usepackage{graphicx}
\usepackage{dcolumn}
\usepackage{bm}
\usepackage{lineno}
\usepackage{tikz}
\usepackage{tikz-feynman}
\usepackage{caption}
\usepackage{amsmath}
\usepackage{float}
\usepackage{subcaption}
\usepackage{xcolor}

\begin{document}
\raggedbottom

\title{ Relic density and astrophysical signatures of MeV scale bound state dark matter}

\author{Pa. Gokhula Prasad}
\email{pa.gokhul1998@gmail.com}
\affiliation{Department of Physics, National Institute of Technology, Tiruchirappalli, India}

\author{V. Suryanarayana Mummidi}
\email{venkata@nitt.edu}
\affiliation{Department of Physics, National Institute of Technology, Tiruchirappalli, India}

\begin{abstract}

We study a strongly interacting complex scalar dark matter candidate, $\chi$, interacting through an attractive potential mediated by a dark radiation $\Phi$. Such interactions allow $\chi$ to form bound states: SIMPonium. We systematically investigate the bound-state dynamics of this dark sector, including the formation and decay of SIMPonium and its impact on the thermal evolution of dark matter. 
We performed Monte Carlo simulations to compute the relic abundances of both free and bound state dark matter. We find that a successful freeze out of the dark sector requires a free dark matter yield of  $Y_\chi \sim 10^{-9}$ and  SIMPonium yield of $Y_B \sim 10^{-13}$. The constraints from $\Delta N_{\rm eff}$ reveal that only scalar mediated self interactions remain viable. For Higgs portal models, these constraints require the dark matter mass to satisfy $m_\chi < 25~\mathrm{MeV}$. Within this mass range, indirect detection constraints imply that, for $g_\chi > 10^{-2}$, the allowed range of the Higgs portal coupling is approximately $10^{-4} \lesssim \lambda \lesssim 10^{-2}$. For $g_\chi < 10^{-2}$, the constraints are satisfied for a broader range, $10^{-4} \lesssim \lambda \lesssim 1$.

\end{abstract}

\maketitle

\tableofcontents

\section{Introduction}

The Planck satellite determined the dark matter abundance  to be 
$\Omega h^2_{DM} = 0.12$, indicating that dark matter constitutes nearly 25\% of 
the present Universe \cite{planck}. The leading cold dark matter candidate known as  Weakly Interacting Massive Particles (WIMP), 
naturally reproduces this relic abundance through the  WIMP miracle \cite{WM}. This theoretical appeal motivated extensive direct detection \cite{DD1,DD2,DD3}, collider searches \cite{DD4,DD5,DD6} and 
indirect detection \cite{ID2,ID3,ID4,ID5,ID6,ID7}. Despite decades of effort, no conclusive evidence 
for WIMP dark matter has been obtained \cite{Nr1,NR2,NR3,NR4}. The absence of experimental 
confirmation strongly motivates the consideration of alternative dark matter 
scenarios.

This need for alternatives becomes even more pressing when viewed alongside 
the well known small scale challenges of the $\Lambda$CDM model, such as core cusp and too big to fail problems \cite{SC1,SC2}. A 
particularly compelling alternative is self-interacting dark matter 
(SIDM) which would solve the small scale issues \cite{SI1,SI2}.  If these self-interactions are sufficiently strong while the coupling to the 
Standard Model remains extremely weak, the framework naturally extends to the 
Strongly Interacting Massive Particle (SIMP) paradigm \cite{SIMP1,SIMP2}. In this 
scenario, the observed relic abundance is achieved through the ``SIMP miracle,'' 
where number changing $4 \rightarrow 2$ interactions in the dark sector govern 
thermal freeze out of dark matter \cite{SIMP3,md2}.

In this work, we consider a complex scalar field $\chi$ as the dark matter 
candidate self interacting through an attractive Coulomb 
potential mediated by a massless mediator $\Phi$. These interactions 
permit the formation of dark bound states by emitting a mediator, which we refer to as 
SIMPonium $(B_n)$. Previous studies have investigated bound state formation in WIMP dark matter models and non radiative bound state formation in SIMP dark matter \cite{WIMP1,WIMP2,BSF3,SIMP4}. In this work, we analyze the impact due to radiative bound state formation by using the formalism suggested by \cite{BSF3}. We also examine the possibility that bound states may 
form not only in the ground state but also in higher excited states, and we 
specifically study the ground state and the first two excited levels of 
SIMPonium $(B_{n'})$.  

We investigate this framework in the early Universe, starting by studying its production mechanism. Using Monte Carlo simulations, we solve the coupled Boltzmann equations that track the abundances of both free and bound-state dark matter. We also examine how each individual process associated with the production mechanism affects the overall yield. Furthermore, since the model includes an extra dark radiation component, $\Phi$, we compute its contribution to the effective number of neutrino species, $\Delta N_{\text{eff}}$. In addition, we determine the temperature at which the dark sector kinetically decouples from the thermal bath.

Even though the signals arising purely from dark radiation are invisible, our 
dark matter candidate can still produce observable signatures through two key 
mechanisms: (i) annihilation of free $\chi$ particles into Standard Model 
states, and (ii) the decay of SIMPonium bound states into Standard Model 
particles. We compute the resulting spectra from final state radiation \cite{IDe1} as well 
as from radiative decays of the Standard Model particles  produced in these 
processes \cite{IDe2} and constrain the model's parameter space using Cosmic Microwave Background  $s$-wave annihilation limits, $511\text{ keV}$ observations, and Integral X-ray data.

The paper is structured as follows. In Section~\ref{sec:model}, we introduce the complex scalar $\chi$ as dark matter candidate
, derive the interaction potential between two dark matter particles, and obtain the corresponding wavefunctions 
of the resulting potential. In Section~\ref{sec:rd}, we solve the coupled Boltzmann 
equations governing the evolution of the abundances of both free and bound 
dark matter, demonstrating how the model reproduces the observed relic density and also in this section we determine $\Delta N_{eff}$ due to dark radiation and also determine temperature at which dark sector completely decouple from thermal bath.
In Section~\ref{sec:id}, we explore the indirect 
detection prospects of the model by computing the photon spectra arising from 
final state radiation and radiative decays. Finally, Section~\ref{sec:con} summarizes our 
study and presents our main conclusions.
\section{Model}
\label{sec:model}
Bound state formation within the SIMP paradigm requires a dark sector capable of both strong self interactions and efficient number changing process. To capture these features in a minimal model, we consider a complex scalar dark matter candidate $\chi$ described by the Lagrangian,
\begin{equation*}
    \mathcal{L} \supset g_\chi\, \chi^* \Phi \chi \;-\; \frac{\lambda_\chi}{4!}\, |\chi|^4-\frac{\lambda}{4!} \chi \chi^* H^\dagger H.
\end{equation*}
The gauge coupling $g_\chi$ provides the attractive force necessary for SIMPonium formation~\cite{md1}, $\lambda$ is coupling that determine the strength of interaction between dark sector and standard model sector via Higgs portal and the quartic term proportional to $\lambda_\chi$ enables the $4\chi \rightarrow 2\chi$ annihilation responsible for obtaining the correct relic density~\cite{md2,SIMP2}.

\medskip

At lowest order, the attractive force between $\chi^*$ and $\chi$ arises from the exchange of the massless boson $\Phi$, represented by the diagram:
\[
W(a_1,a_2;b_1,b_2) =
\begin{tikzpicture}[baseline=(current bounding box.center)]
  \begin{feynman}
    \vertex (in1)  {\(\chi^*(a_1)\)};
    \vertex (in2) [below=2cm of in1] {\(\chi(a_2)\)};

    \vertex (out1) [right=3cm of in1] {\(\chi^*(b_1)\)};
    \vertex (out2) [below=2cm of out1] {\(\chi(b_2)\)};

    \vertex (ph1) [right=1.75cm of in1];
    \vertex (ph2) [right=1.75cm of in2];

    \diagram*{
      (in1) -- [anti fermion] (out1),
      (in2) -- [fermion] (out2),
      (ph1) -- [photon, edge label'=\(\Phi\)] (ph2)
    };
  \end{feynman}
\end{tikzpicture}
\]

In the non-relativistic limit, the scattering amplitude simplifies to,
\begin{equation*}
    W(a_1 - b_1) = \frac{4 m_\chi^2 g_\chi^2}{(a_1 - b_1)^2}.
\end{equation*}
The corresponding interaction potential is given by \cite{md5},
\begin{equation*}
    V(r) = \frac{i}{4 \mu m_\chi} \int \frac{d^3 p}{(2\pi)^3} W(p) e^{-i p \cdot r},
\end{equation*}
which results in an attractive Coulomb potential,
\begin{equation*}
    V(r) = -\frac{V_0}{r}, \quad V_0 > 0,
\end{equation*}
where $V_0 = \frac{g_\chi^2}{4 \pi}$ is the interaction strength and $\mu = \frac{m_\chi}{2}$ is the reduced mass.

\medskip

The Coulomb potential gives two classes of eigenstates: discrete bound levels and a continuous energy spectrum. The continuous spectrum corresponds to scattering states of free dark matter particles and is characterized by a continuous quantum number associated with the expectation value of the relative momentum, $P = \mu v_{\rm rel}$. In contrast, the discrete spectrum represents the bound states of dark matter, defined by discrete \textit{nlm} quantum numbers. We extend the analysis up to $n=3$ to examine the properties of the corresponding excited states.

\medskip

The bound state $\Psi(r)_n$ and scattering state $\Psi(r)_s$ of SIMPonium under this potential are \cite{md4},
\begin{equation}
    \Psi(r)_n = N_{nlm} \, r^l \, e^{-\sqrt{\alpha} r} \, L_n^{2l+1}(2 r \sqrt{\alpha}) \, Y^l_m(\theta, \phi),
\end{equation}
\begin{equation}
    \Psi(r)_s = N \, e^{i P \cdot r} \, {}_1F_1(-i\delta; 1; i (P r - \mathbf{P}\cdot\mathbf{r})).
\end{equation}
Here, $L_n^{2l+1}(2 r \sqrt{\alpha})$ is the associated Laguerre polynomial, and ${}_1F_1(-i\delta; 1; i (P r - \mathbf{P}\cdot\mathbf{r}))$ is the confluent hypergeometric function of the first kind. A detailed derivation is presented in Appendix~\ref{app:A}.

In our framework, free dark matter particles can annihilate into Standard Model  particles through the Higgs portal, owing to their interaction with the Higgs boson. The corresponding annihilation cross section is given by
\begin{equation}
(\sigma v){ Ann} =
\frac{2\pi\zeta}{1-e^{-2\pi\zeta}}
\frac{|M|^2}{8\pi s}
\frac{\sqrt{s-4m{\rm SM}^2}}{\sqrt{s}},
\label{cs_ann}
\end{equation}
where $s$ is the Mandelstam variable, and $\zeta=V_0/v_{\rm rel}$, with $V_0$ characterizing the strength of the Coulomb potential and $v_{\rm rel}$ denoting the relative velocity of the annihilating dark matter particles. The prefactor accounts for the Sommerfeld enhancement arising from the long-range interaction between the incoming dark matter particles.

For annihilation into a pair of SM fermions, the squared perturbative matrix element is given by
\begin{equation*}
|M|^2 =
2(s-m_f^2)
\left(
\frac{\lambda m_f}{s-m_h^2}
\right)^2,
\end{equation*}
where $\lambda$ denotes the Higgs-portal coupling, $m_f$ is the fermion mass, and $m_h$ is the Higgs-boson mass. Since we focus on the low-energy regime with $m_\chi=\mathcal{O}(\mathrm{MeV})$, annihilation into hadronic final states is described using Chiral Perturbation Theory ($\chi$PT). In particular, the squared perturbative matrix element for annihilation into a pair of pions is given by
\begin{equation*}
|M|^2 =
\left(
\frac{\lambda m_\pi^2}{s-m_h^2}
\right)^2,
\end{equation*}
where $m_\pi$ is the pion mass. Substituting the corresponding matrix elements into Eq.~\eqref{cs_ann}, we obtain the annihilation cross sections for the fermionic and pionic final states.

In addition to the annihilation of free dark matter particles, the formation of SIMPonium bound states introduces another mechanism for producing SM particles. Since SIMPonium consists of a dark matter particle and its antiparticle, it can annihilate into SM final states through the Higgs portal. Consequently, these bound states are unstable and can decay into kinematically accessible SM particles. The corresponding decay rate is given by,
  \begin{equation}
  \Gamma_{B_n} = (\sigma v)_0 \, |\psi_n(0)|^2,
  \label{cs_de}
\end{equation}
where $(\sigma v)_0 $ is perturbative cross section.
\section{SIMPonium and free $\chi$ during the early universe}
\label{sec:rd}
\begin{figure}[t]
  \centering
  \begin{tikzpicture}[font=\small]
  \begin{feynman}
    \vertex (blob) at (0,0) [draw, circle, minimum size=1cm, fill=gray!30] {};

    \vertex (in1) [above left=2cm and 2cm of blob] {\(\chi^*\)};
    \vertex (in2) [below left=2cm and 2cm of blob] {\(\chi^*\)};
    \vertex (in3) [above left=1cm and 2cm of blob] {\(\chi\)};
    \vertex (in4) [below left=1cm and 2cm of blob] {\(\chi\)};

    \vertex (out1) [above right=1cm and 2cm of blob] {\(\chi^*\)};
    \vertex (out2) [below right=1cm and 2cm of blob] {\(\chi\)};

    \diagram*{
      (in1) -- [anti fermion] (blob) -- [anti fermion] (out1),
      (in2) -- [anti fermion] (blob) -- [fermion] (out2),
      (in3) -- [fermion] (blob),
      (in4) -- [fermion] (blob)
    };
  \end{feynman}
\end{tikzpicture}
  \caption{Feynman diagram illustrating the $4\chi \rightarrow 2\chi$ number changing process.}
  \label{feyn_42}
\end{figure}
We assume that in the early Universe, dark matter  is in thermal equilibrium with the Standard Model thermal bath through both chemical and kinetic equilibrium. Chemical equilibrium is maintained as long as the rate of number changing interactions exceeds the Hubble expansion rate. Similarly, kinetic equilibrium is maintained when the scattering rate exceeds the Hubble expansion rate. Many  dark matter models such as WIMPs, maintains chemical equilibrium with the Standard Model  bath in the early Universe through annihilation processes of the form
\(\chi \chi^\ast \rightarrow \mathrm{SM}\,\overline{\mathrm{SM}}\)~\cite{rd3,rd4}. Unlike WIMPs, SIMP dark matter particles follow the inequality relation \cite{SIMP3},
\begin{equation}
\Gamma_{\chi + {\rm SM} \to \chi + {\rm SM}}
\;\gtrsim\;
\Gamma_{4\chi \to 2\chi}
\;\gg\;
\Gamma_{2\chi \to 2\,{\rm SM}} \, , 
\label{eq:ineq}
\end{equation}
where $\Gamma_{\chi + {\rm SM} \to \chi + {\rm SM}} = n_{SM} \langle \sigma v \rangle$, $\Gamma_{4\chi \to 2\chi} = n_\chi^3 \langle \sigma v^3 \rangle$ and $\Gamma_{2\chi \to 2\,{\rm SM}} = n_\chi \langle \sigma v \rangle$.
The condition imposed by Eq.~\eqref{eq:ineq} suggests that the interaction rate for dark matter annihilation into Standard Model particles (\(\chi \chi^\ast \rightarrow \mathrm{SM}\,\overline{\mathrm{SM}}\))  is feeble and number density of $\chi$ changes primarily due to self interactions  such as
\(4\chi \rightarrow 2\chi\). Despite the  feeble coupling between the dark and visible sectors, the large number density of Standard Model particles enhances the $\chi\,\mathrm{SM} \rightarrow \chi\,\mathrm{SM}$
scattering rate sufficient enough  to maintain thermal equilibrium between the two sectors.

\begin{table}[b]
    \caption{Relevant SIMP and bound state processes contributing to the relic abundance, along with their thermally averaged cross sections in natural units.}
    \label{42process}
    \begin{ruledtabular}
    \begin{tabular}{clc}
        Label & Process & Benchmark   \\
        \colrule
        \noalign{\vskip 3pt}
        A & $\chi + \chi^* \rightarrow B_n + \Phi$
          & $8.54 \times 10^{-25}$ \\
        B & $B_n + \Phi \rightarrow \chi + \chi^*$
          & $ 1.68 \times 10^{-5}$ \\
        C & $\chi+\chi+\chi^*+\chi^* \rightarrow \chi^*+\chi$
          & $1.35 \times 10^{-7}$ \\
        D & $\chi+\chi+\chi^*+\chi^* \rightarrow B_n + \Phi$
          & $6.8 \times 10^{-16}$ \\
        E & $\chi+\chi^*+B_n \rightarrow \chi+\chi^*$
          & 0.03651  \\
        F & $B_n+B_n \rightarrow \chi+\chi^*$
          & $4.55 \times 10^{-26}$ \\
        G & $\chi+\chi^*+B_n \rightarrow B_n + \Phi$
          & 0.2237 \\
        H & $B_n+B_n \rightarrow B_n + \Phi$
          & $1.2 \times 10^{-8}$ \\
        I & $B_n \rightarrow \Phi + \Phi$
          & $3.02 \times 10^{-18}$ \\
        J & $\chi+\chi^* \rightarrow \Phi + \Phi$
          & $1.4 \times 10^{-30}$ \\
    \end{tabular}
    \end{ruledtabular}
\end{table}

\subsection{Relic density of free and bound state dark matter}

In this section, we study the evolution of the number densities of free dark matter $(\chi)$ and SIMPonium $(B_n)$, and also determine the temperature at which dark matter departs from chemical equilibrium. In scenarios where bound states do not form, the thermal evolution of $\chi$ would be determined by a simplest $4\chi \rightarrow 2\chi$ interaction, $\chi\,\chi\,\chi^*\,\chi^* \rightarrow \chi^* \, \chi$
. On the flip side, the scenarios in which bound states could form, new particles such as   SIMPonium ground state $B_n$ and its excited state $B_{n'}$ would coexist along with free dark matter $\chi$. The presence of $B_n$ and $B_{n'}$ would modify the evolution of dark matter abundance by introducing new $4\chi \rightarrow 2\chi$ annihilation process involving bound states. All such processes relevant for the Boltzmann evolution are listed in Table~\ref{42process} and the following discussion highlights key features of these interactions. Following the bound state  formation through channels A and D, the   bound states undergo further evolution via the processes labeled B, E, F, G, H, and I . Among these, only process G preserves the bound state number density. In contrast, processes B, E, F,H and I  modify the bound state number density, where B and I corresponds to SIMPonium ionization and decay respectively.

The thermal evolution of the dark matter particle $\chi$ and the bound state $B_n$ is governed by a set of coupled Boltzmann equations written in terms of the yield variables
$Y_\chi = \frac{n_\chi}{s} $ and $Y_B = \frac{n_B}{s}$, where $n_\chi$ and $n_B$ denote the number densities of the free dark matter particles and the bound states, respectively, and $s$ is the entropy density of the Universe. 
It is convenient to parametrize the evolution using the dimensionless variable
$x = \frac{m_\chi}{T}$, where $m_\chi$ is the dark matter mass and $T$ denotes the temperature of the Standard Model thermal bath.
 The Boltzmann equations describing the evolution of the dark matter and SIMPonium abundances are given in Eq.~\eqref{eq:dYchi} and \eqref{eq:dYBn}. The right-hand sides of these equations contain the relevant number changing collision terms, involving $\langle \sigma v \rangle$, $\langle \sigma v^2 \rangle$, $\langle \sigma v^3 \rangle$ and the analytical expressions for these cross sections are derived in detail in Appendix~\ref{app:B}.

\begin{widetext}

\begin{align}
\frac{dY_{\chi}}{dx} &= \frac{x}{s H(m)} \Bigg[
   - s^2 \langle \sigma v \rangle_A
     \left( Y_\chi^2 - \frac{(Y_\chi^{\rm eq})^2 \, Y_{B_n} }{Y_{B_n}^{\rm eq} } \right)
+ s^2 \langle \sigma v \rangle_B 
     \left( Y_{\Phi}^{eq} \, Y_{B_n} - \frac{Y_\chi^2 \, Y_{B_n}^{\rm eq} \, Y_{\Phi}^{\rm eq}}{(Y_\chi^{\rm eq})^2} \right)  - s^4 \langle \sigma v^3 \rangle_C 
     \left( Y_\chi^4 - (Y_\chi^{\rm eq})^2 \, Y_\chi^2 \right) \nonumber \\
&\quad - s^4 \langle \sigma v^3 \rangle_D
     \left( Y_\chi^4 - \frac{(Y_\chi^{\rm eq})^4 \, Y_{B_n} }{ Y_{B_n}^{\rm eq} } \right) 
 + s^2 \langle \sigma v \rangle_F 
     \left( Y_{B_n}^2 - \frac{(Y_{B_n}^{\rm eq})^2 Y_\chi^2}{(Y_\chi^{\rm eq})^2} \right)  
- s^3 \langle \sigma v^2 \rangle_G 
     \left( Y_{B_n} Y_{\chi}^2 - (Y_\chi^{\rm eq})^2 Y_{B_n}  \right) \nonumber \\
 &\quad   - s^ 2\langle \sigma v \rangle_K 
     \left( Y_{\chi}^2 - (Y_\chi^{\rm eq})^2   \right)
\Bigg],
\label{eq:dYchi}
\end{align}

\begin{align}
\frac{dY_{B_n}}{dx} &= \frac{x}{s\, H(m)} \Bigg[
    s^2 \langle \sigma v \rangle_A
     \Bigg( Y_\chi^2 - \frac{(Y_\chi^{\rm eq})^2 \, Y_{B_n} }{Y_{B_n}^{\rm eq} } \Bigg)
     - s^2 \langle \sigma v \rangle_B
     \Bigg( Y_{\Phi}^{eq} \, Y_{B_n} - \frac{Y_\chi^2 \, Y_{B_n}^{\rm eq} \, Y_{\Phi}^{\rm eq}}{(Y_\chi^{\rm eq})^2} \Bigg) 
  + s^4 \langle \sigma v^3 \rangle_D
     \Bigg( Y_\chi^4 - \frac{(Y_\chi^{\rm eq})^4 \, Y_{B_n} }{ Y_{B_n}^{\rm eq} } \Bigg) \nonumber \\
    &\quad   - s^3 \langle \sigma v^2 \rangle_E
     \Bigg( Y_\chi^2 Y_{B_n} - (Y_\chi^{\rm eq})^2 Y_{B_n}^{\rm eq} \Bigg) 
 - s^2 \langle \sigma v \rangle_F 
     \Bigg( Y_{B_n}^2 - \frac{(Y_{B_n}^{\rm eq})^2 \, Y_\chi^2}{(Y_\chi^{\rm eq})^2} \Bigg) 
      - s^2 \langle \sigma v \rangle_H
     \Bigg( Y_{B_n}^2 - Y_{B_n}^{\rm eq} \, Y_{B_n}  \Bigg)  \nonumber \\
 &\quad - s \langle \Gamma_I \rangle  \Big( Y_{B_n} - Y_{B_n}^{\rm eq}  \Big)
\Bigg].
\label{eq:dYBn}
\end{align}
\end{widetext}

Since the dark radiation $\Phi$ is a 
relativistic particle, we enforce 
$Y_{\Phi} = Y_{\Phi}^{\rm eq}$ and  simplifying the coupled 
Boltzmann equations accordingly. We also study the evolution of $\chi$  in the absence of bound-state formation by solving the corresponding Boltzmann equation:

\begin{equation}
  \frac{dY_{free}}{dx} = \frac{-x}{ H(m)}  s^3 \langle \sigma v^3 \rangle_C
     \left( Y_\chi^4 - (Y_\chi^{\rm eq})^2 \, Y_\chi^2 \right).
     \label{eq:yfree}
\end{equation}
\begin{figure}[!htbp]
  \centering

  \begin{minipage}{0.9\linewidth}
        \centering
        \text{(a)}\\[2mm]
        \includegraphics[width=\linewidth]{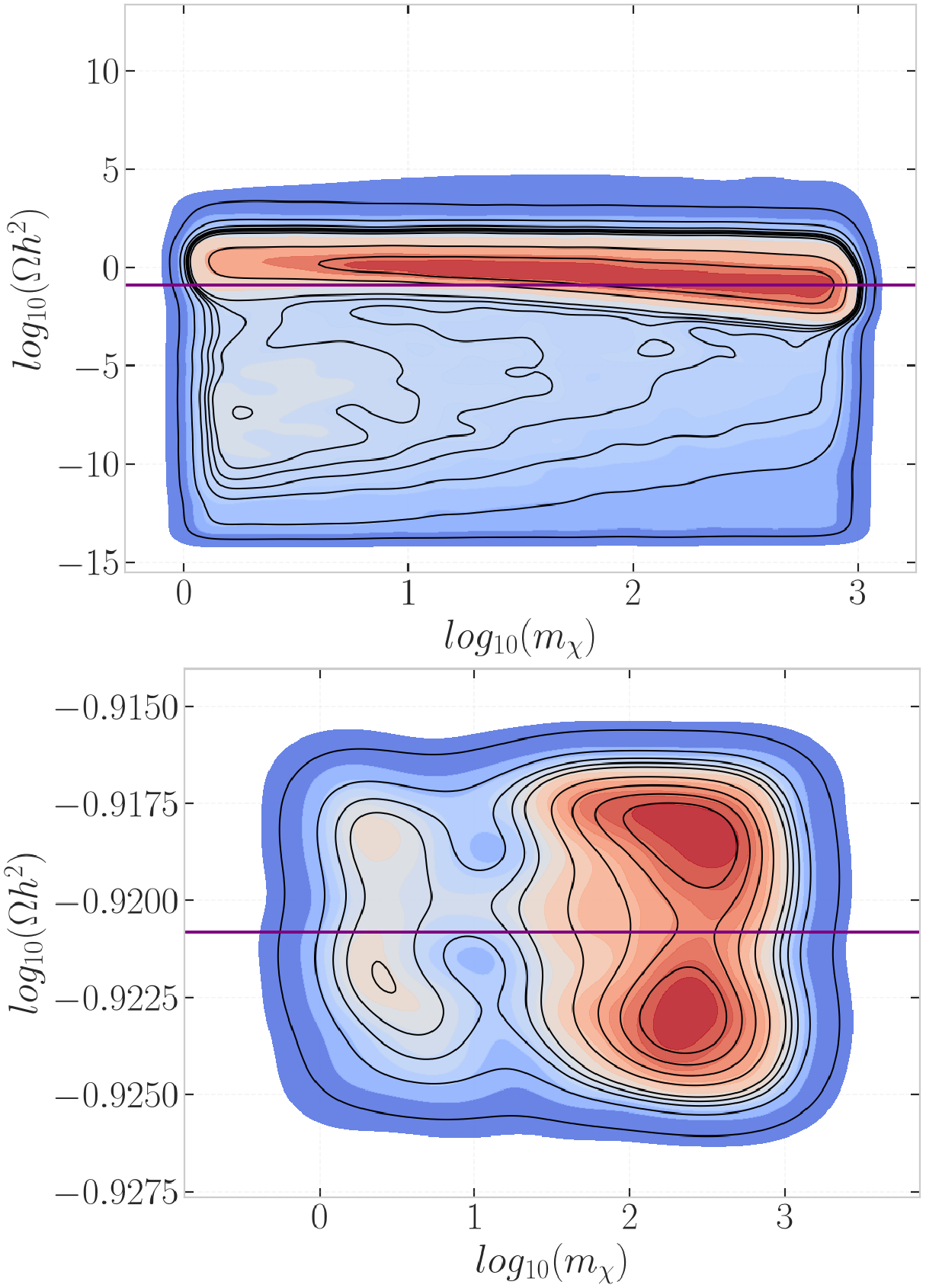}\\[2mm]
        \text{(b)}
    \end{minipage}

  \caption{Kernel density estimation  of the relic abundance obtained from Monte Carlo simulations. Panel (a) shows the distribution of all simulated points, while panel (b) shows the distribution of points satisfying the observed relic-density constraint. The purple horizontal line indicates the central observed value, $\Omega h^2 = 0.12$.}
  \label{plot_rd}
\end{figure}

\begin{figure}[!htbp]
  \centering
 \begin{minipage}{0.9\linewidth}
        \centering
        \text{(a)}\\[2mm]
        \includegraphics[width=\linewidth]{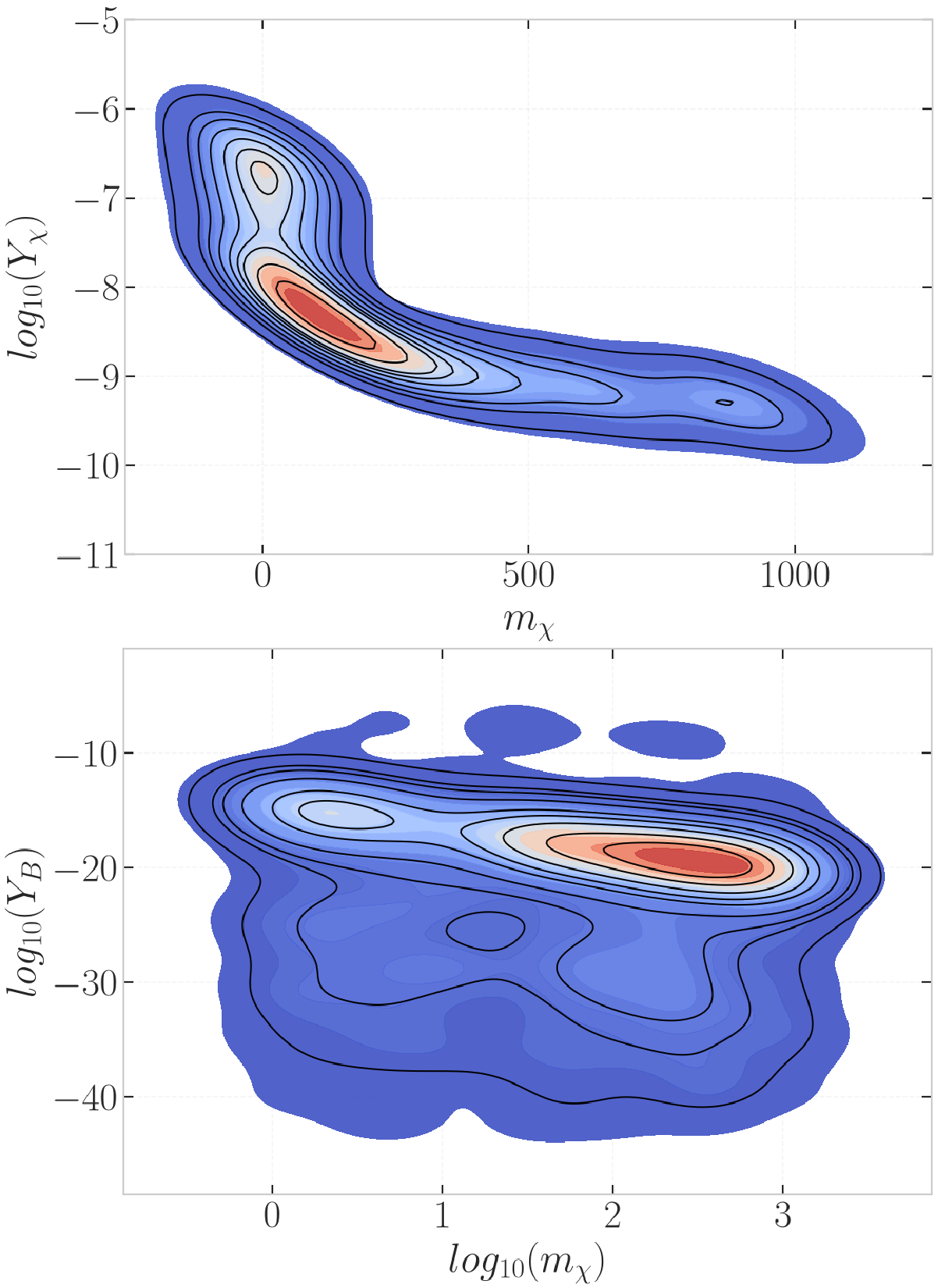}\\[2mm]
        \text{(b)}
    \end{minipage}

  \caption{KDE plots of the yields of dark matter $Y_\chi$ and SIMPonium $Y_B$ for parameter points that reproduce the observed dark matter relic abundance.}
  \label{plot_y}
\end{figure}

\begin{figure*}[t]
    \centering
    \includegraphics[width=\textwidth]{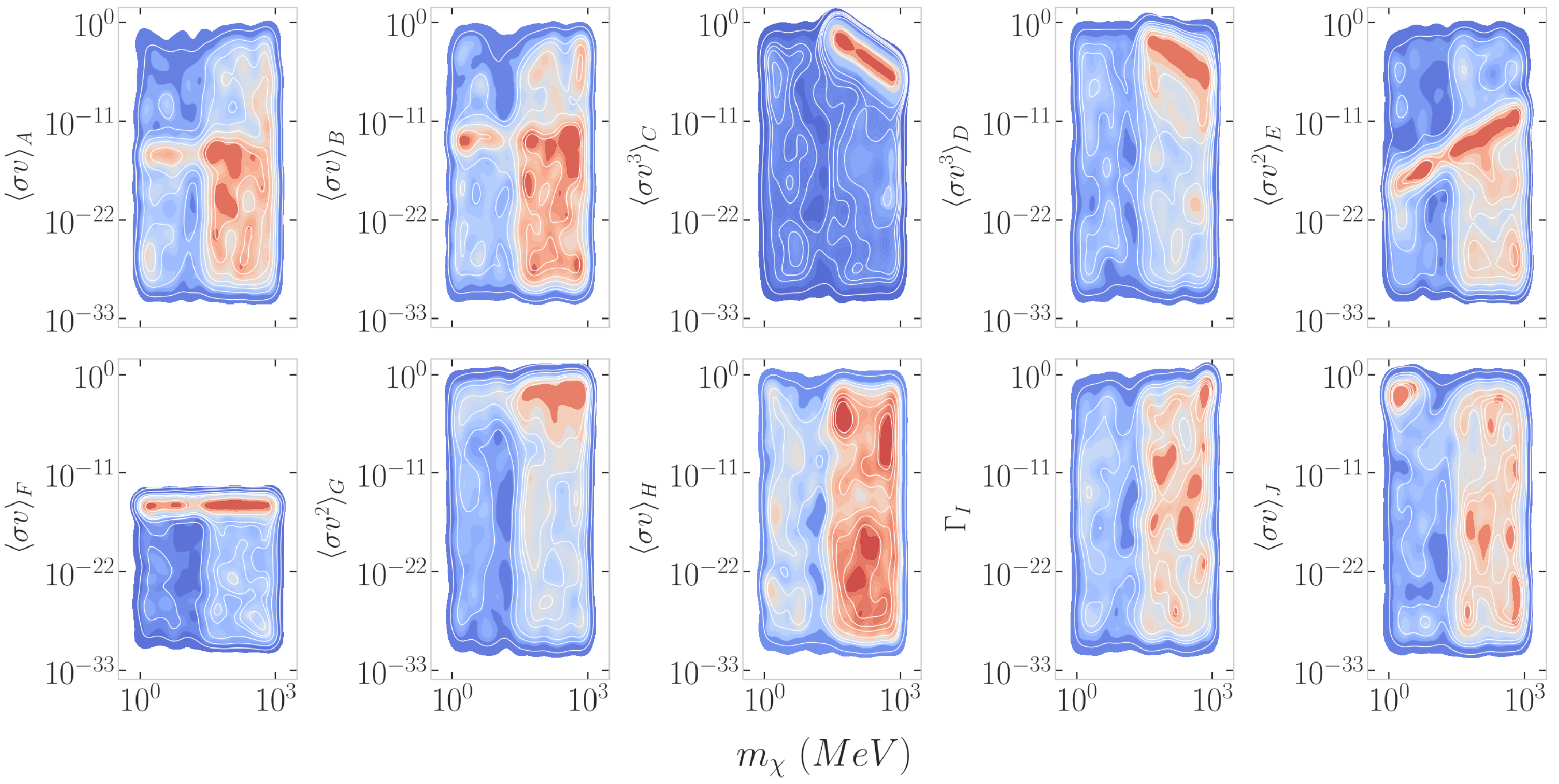}
    \caption{KDE distributions of the thermally averaged cross sections and interaction rates for all processes relevant to determining the dark matter relic abundance.}
    \label{plot_TCS}
\end{figure*}

In this work, we solve the coupled Boltzmann equations in a model independent manner using a Monte Carlo scan over the relevant parameter space. Specifically, we randomly sample the thermally averaged cross sections, $\langle \sigma v^n \rangle_i \in [10^{-30},10^{0}]$, and the dark matter mass, $m_\chi \in [10^{0},10^{3}]\,\mathrm{MeV}$, and use these samples as inputs to the coupled system of equations to determine the resulting relic abundance.

Figure~\ref{plot_rd} presents the kernel density estimation (KDE) of the relic abundance obtained from our Monte Carlo simulations. In panel (a), the KDE distribution of the complete simulated dataset illustrates how the sampled parameter points populate the $(m_\chi,\Omega h^2)$ plane. A substantial concentration of points is observed in the vicinity of the measured relic abundance, $\Omega h^2 \simeq 0.12$, indicating that the explored parameter space contains numerous configurations yielding relic abundances close to the observed value. To investigate this further, panel (b) shows the KDE distribution of the subset of points satisfying $\Omega h^2 = 0.12 \pm 0.0012$. The resulting distribution exhibits a higher concentration of relic density compatible points for dark matter masses $m_\chi \gtrsim \mathcal{O}(10^1)\,\mathrm{MeV}$. 

Figure~\ref{plot_y} shows the KDE distributions of the dark matter yield, \(Y_\chi\), and the SIMPonium yield, \(Y_B\), for parameter points that reproduce the observed relic abundance, \(\Omega_{\rm DM}h^2\). The allowed values of \(Y_\chi\) span approximately \(10^{-10}\) to \(10^{-6}\), with most parameter points concentrated around \(Y_\chi\sim10^{-9}\). In contrast, \(Y_B\) spans a much wider range, from approximately \(10^{-40}\) to \(10^{-10}\), with the highest concentration of points around \(Y_B\sim10^{-13}\). This broad range of \(Y_B\) values, even for a fixed \(Y_\chi\), indicates that the model can accommodate substantially different abundances of the bound state \(B\) while satisfying the observed relic abundance. Consequently, the model allows for scenarios in which the stable dark matter component \(\chi\) undergoes self-annihilation, while a fraction of the dark matter abundance is stored in SIMPonium bound states \(B\) that subsequently decay. The relative abundance of this decaying bound-state component can vary considerably, ranging from a highly subdominant fraction to a potentially significant contribution relative to the stable dark matter component, while remaining consistent with the observed relic abundance. 

Figure~\ref{plot_TCS} shows the KDE distributions of the thermally averaged cross sections and interaction rates associated with the processes included in the coupled Boltzmann equations, for parameter points that reproduce the observed dark matter relic abundance. These distributions illustrate how different interaction strengths can contribute to reproducing the observed relic abundance within the coupled system. The thermally averaged cross sections for bound state formation and ionization are predominantly concentrated within the range $10^{-30}-–10^{-11}$. In contrast, the  processes associated with cannibalistic behavior, namely $\langle\sigma v^3\rangle_C$, $\langle\sigma v^3\rangle_D$, and $\langle\sigma v^2\rangle_G$, generally require larger cross sections, spanning approximately $10^{-5}$ to $\mathcal{O}(10^0)$. Notably, $\langle\sigma v\rangle_E$ is the only process for which all accepted parameter points lie below $10^{-11}$, whereas the other processes exhibit values both below and above this threshold. Furthermore, the annihilation of dark matter into pairs of mediators and the SIMPonium decay rate exhibit broad distributions across the sampled parameter space, spanning approximately $10^{-30}$ to $10^0$. This wide variation indicates that the observed relic abundance can be achieved across a broad range of self interaction strengths.

\begin{figure}[t]
  \centering
 
  \includegraphics[width=\linewidth]{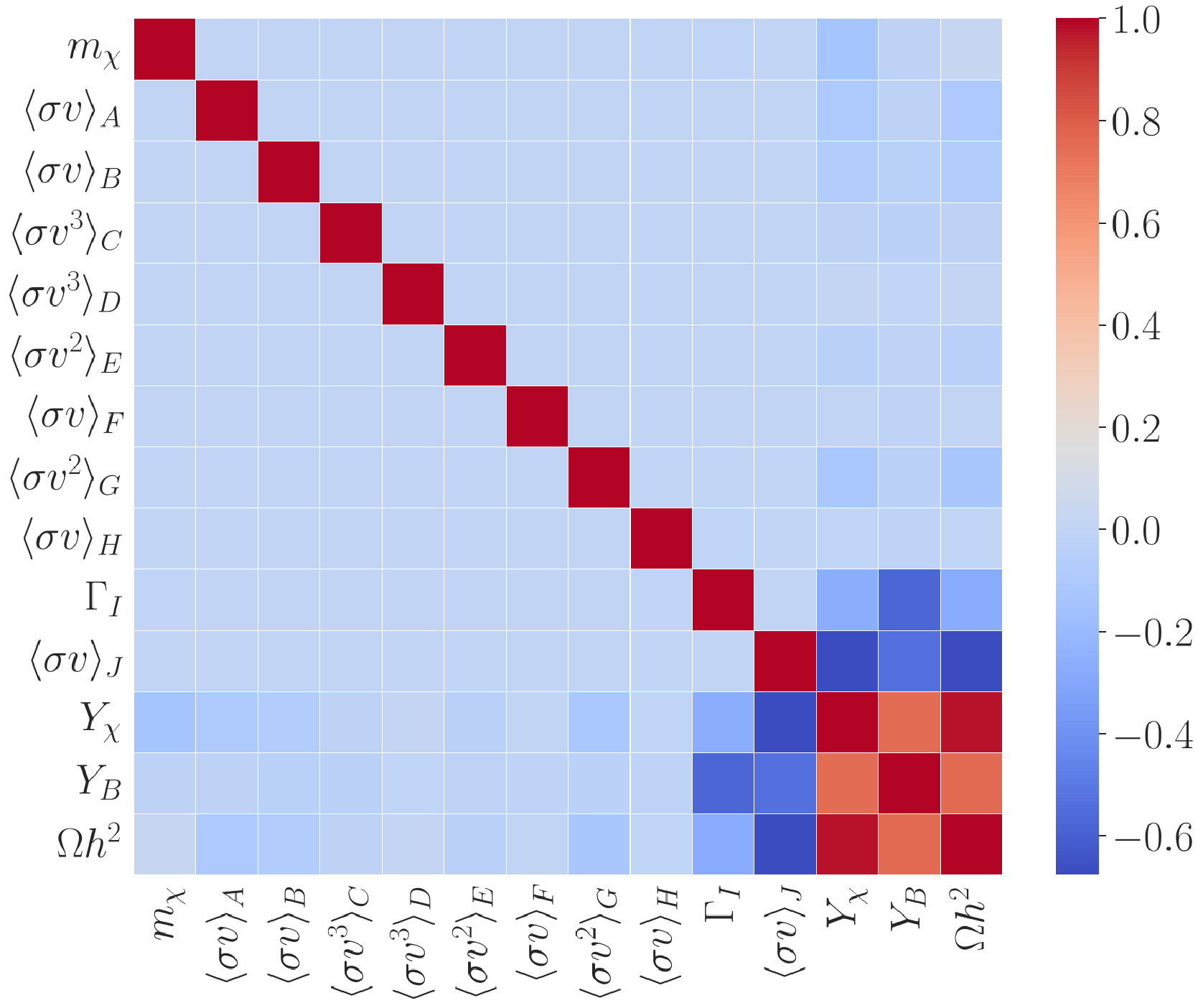}

  \caption{Corrolation heatmap of mass of dark matter, cross section and relic density}
  \label{plot_corr}
\end{figure}

Figure~\ref{plot_corr} presents the correlation matrix between the dark matter mass $m_\chi$, all relevant interaction processes, the final yields, and the relic abundance. Although the model parameters and interaction rates exhibit varying degrees of correlation with the final yields and relic abundance, dark matter self annihilation into mediator pairs and SIMPonium decay into mediator pairs show particularly strong negative correlations with these quantities. This indicates that these two processes play a crucial role in determining the final dark matter abundance and achieving the observed relic density. 

\begin{figure}[!htbp]
  \centering
 
  \includegraphics[width=\linewidth]{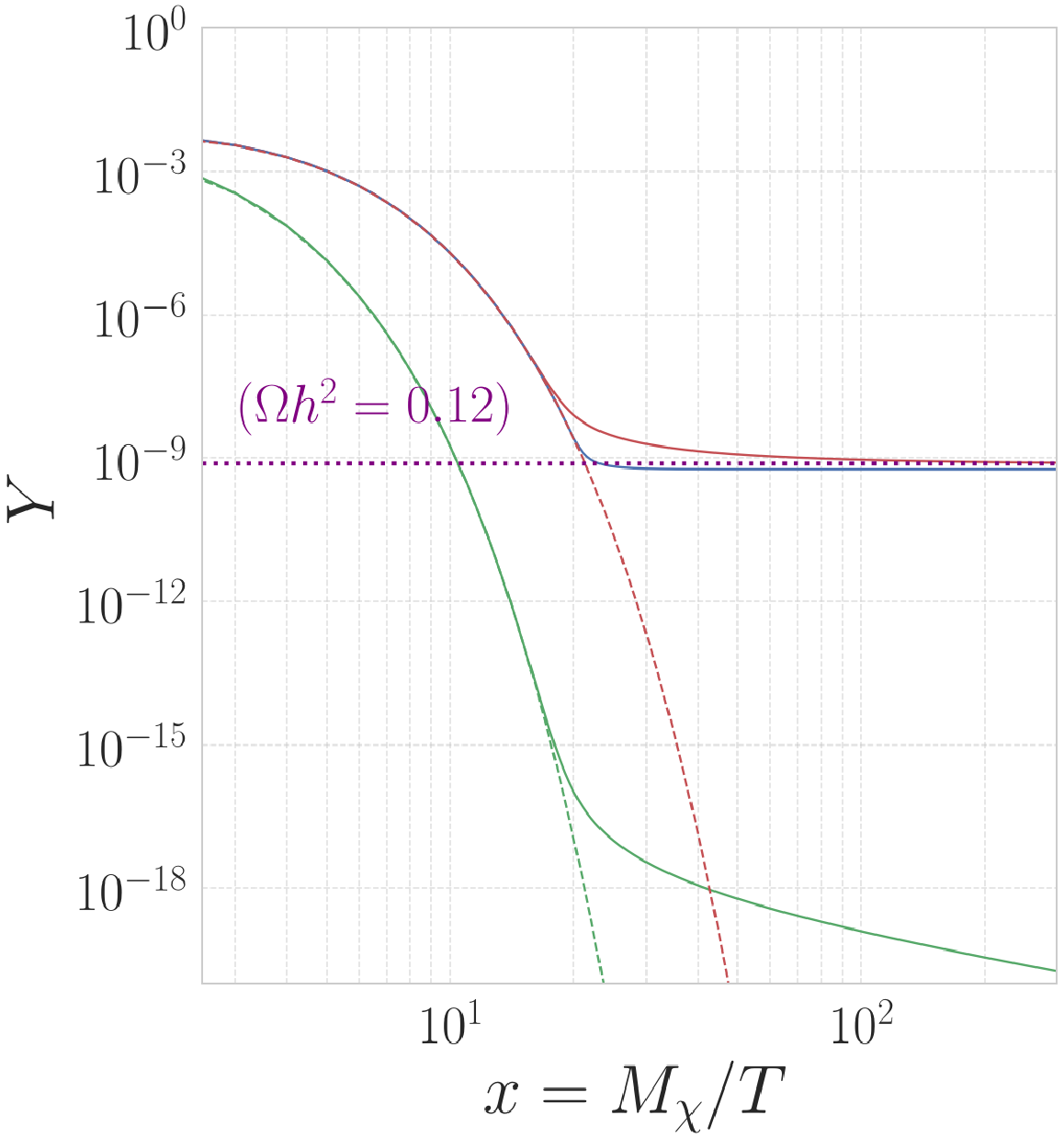}

  \caption{Evolution of the number densities of free dark matter (red), SIMPonium (green), and the dark matter abundance in the absence of bound state formation (blue). The solid curves represent the evolution of the abundances, while the corresponding dotted curves denote their equilibrium abundances. The observed relic density $\Omega h^2 = 0.12 \pm 0.0012$ is indicated by the purple dotted line.}
  \label{plot_Yield}
\end{figure}

In Fig.~\ref{plot_Yield}, we present the numerical solutions for $Y_\chi$, $Y_B$, and $Y_{free}$, obtained by solving Eqs.~\eqref{eq:dYchi}, \eqref{eq:dYBn}, and \eqref{eq:yfree}, respectively, using the benchmark cross sections listed in Table~\ref{42process}. In the absence of bound states, the evolution of dark matter is governed solely by $\langle \sigma v^3 \rangle_C$, and $Y_\chi$ tracks its equilibrium value, $Y_\chi^{\mathrm{eq}}$, until chemical decoupling occurs at $x_1 \approx 16$.

However, when bound states are included, process D becomes kinematically allowed and, being slightly weaker than process C, freezes out earlier. Even after process C freezes out, the dark sector does not become completely decoupled, as the processes $\langle \sigma v \rangle_B$, $\langle \sigma v^2 \rangle_E$, and $\langle \sigma v^2 \rangle_G$ remain relevant during this epoch, governing the subsequent evolution and final abundances of $B_n$ and $\chi$. Since $E_n > T$, newly formed bound states undergo rapid ionization, while their out-of-equilibrium decay further depletes their population, driving the bound-state abundance to remarkably low values.
In contrast to non radiative bound state formation, where the SIMPonium abundance freeze out around $x \approx 110$~\cite{SIMP4}, our inclusion of SIMPonium decay ($\Gamma_I$) ensures that standard freeze out is entirely prevented, even when ionization becomes inefficient at temperatures below the binding energy.

The total relic density of dark matter is   obtained as \cite{EU},

\begin{align}
\Omega h^2
 &= 2.752 \times 10^5 \left( \Omega_\chi h^2 + \Omega_B h^2 \right) \nonumber\\
 &= 2.752 \times 10^5
    \Biggl[
      \left( \frac{m_\chi}{\mathrm{MeV}} \right)
      Y_\chi(x \rightarrow \infty)
      \nonumber\\
 &\qquad
      + \left( \frac{m_B}{\mathrm{MeV}} \right)
      Y_B(x \rightarrow \infty)
    \Biggr].
\label{RD}
\end{align}
Since the bound states evolves  without freezing out, for $x\rightarrow \infty$ their final yield becomes negligible compared to that of $\chi$. Consequently, Eq.~\eqref{RD} can be approximated as,
\begin{align}
\Omega h^2
&\approx 2.752 \times 10^5 \left( \Omega_\chi h^2 \right) \nonumber \\
&\approx 2.752 \times 10^5
\left[
\left( \frac{m_\chi}{\mathrm{MeV}} \right)
Y_\chi(x \rightarrow \infty)
\right].
\label{RDapprox}
\end{align}

\subsection{Constraints on $\Delta N_{eff}$}
We assume the presence of a dark mediator that mediates the attractive interaction between dark matter particles and  the presence of dark radiation increases the radiation energy density and can therefore be parametrized in terms of an effective number of  neutrino species. Here, $\Delta N_{\mathrm{eff}}$, arise due to entropy dump from dark sector annihilation or decay into Standard Model  particles and dark radiation. This contribution depends on whether entropy transfer from the dark sector occurs before or after neutrino decoupling, since this determines how the injected energy is shared between the neutrino and photon thermal bath. We distinguish two cases: (i) decoupling of the dark sector occurs before neutrino decoupling, $(T_{\chi\text{-dec}}>T_{\nu\text{-dec}})$, and (ii) decoupling occurs after neutrino decoupling, $(T_{\chi\text{-dec}}<T_{\nu\text{-dec}})$.

Following \cite{ne} the dark sector decouples before neutrino decoupling, $\Delta N_{eff}$ is given by
\begin{equation}
  \Delta N_{eff} (t< t_{\nu - dec})= \frac{4 g_\Phi}{7} \left(\frac{\epsilon}{\eta} \frac{43}{4 g^*_{SM}(T_{\chi-dec})}\right)^\frac{4}{3}.
\end{equation}
where 43/4 is the effective number of   degrees of freedom in the Standard Model thermal bath prior to neutrino decoupling.
If instead the dark sector decouples after neutrino decoupling, $\Delta N_{eff}$ is given by
\begin{equation}
  \Delta N_{eff} (t> t_{\nu - dec})= \frac{4 g_\Phi}{7} \left(\frac{\epsilon}{\eta} \frac{11}{2 g^*_{SM}(T_{\chi-dec})}\right)^\frac{4}{3}.
\end{equation}
Here, the factor 11/2 accounts for the  degrees of freedom in the electromagnetic bath after neutrino decoupling but before electron-positron annihilation. And for scalar mediator $g_\Phi=1$ , for massless vector meditor $g_\Phi=2$
\begin{figure}[!htbp]
  \centering

  \begin{minipage}{0.9\linewidth}
        \centering
        \text{(a)}\\[2mm]
        \includegraphics[width=\linewidth]{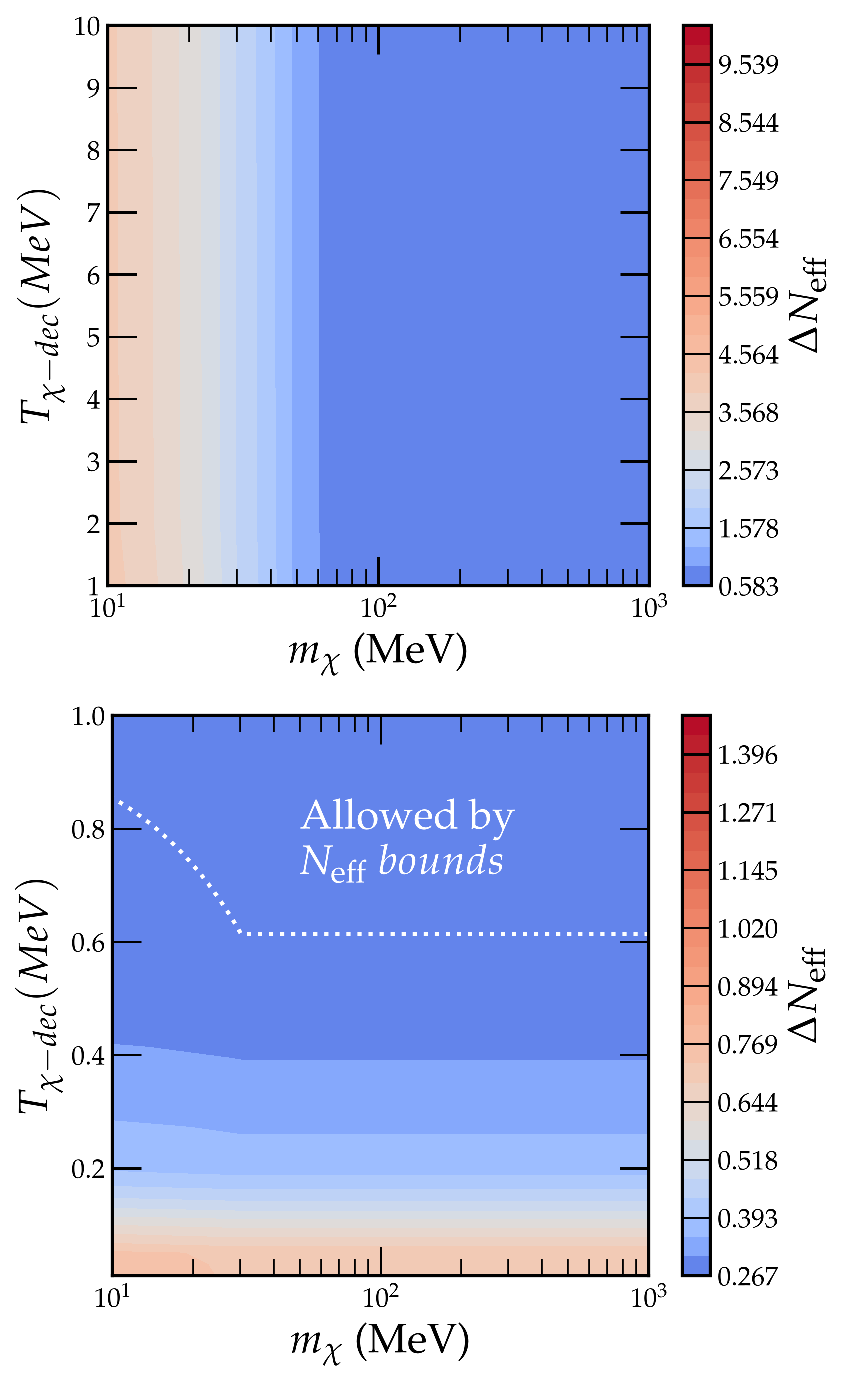}\\[2mm]
        \text{(b)}
    \end{minipage}

  \caption{Effective number of relativistic species, $N_{\mathrm{eff}}$, in the scalar mediated scenario. Panel (a) shows the results for dark sector kinetic decoupling before neutrino decoupling, while panel (b) shows the corresponding results for dark sector decoupling after neutrino decoupling. }
  \label{plot_s}
\end{figure}
\begin{figure}[!htbp]
  \centering

  \begin{minipage}{0.9\linewidth}
        \centering
        \text{(a)}\\[2mm]
        \includegraphics[width=\linewidth]{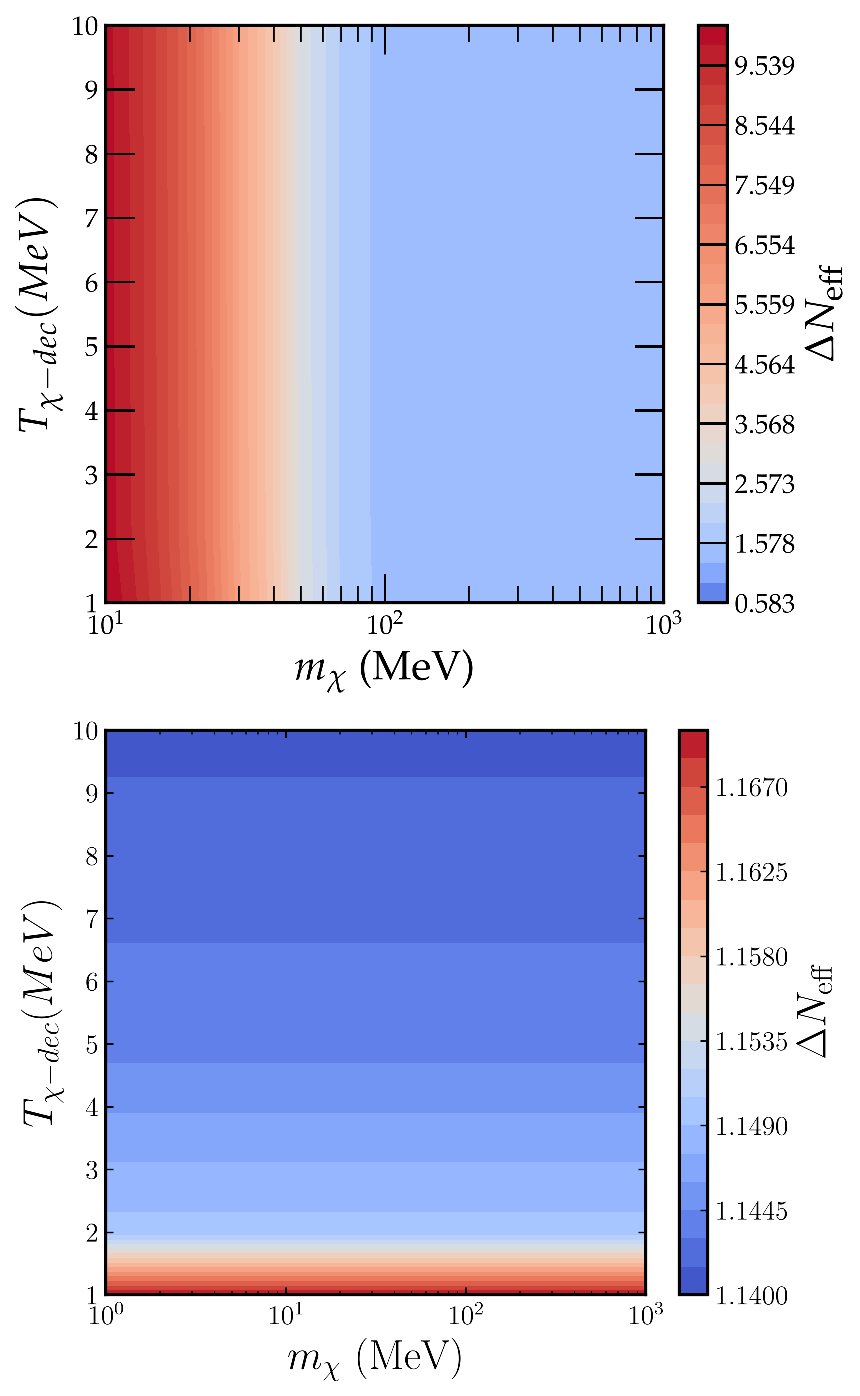}\\[2mm]
        \text{(b)}
    \end{minipage}

 \caption{Effective number of relativistic species, $N_{\mathrm{eff}}$, in the vector mediated scenario. Panel (a) shows the results for dark sector kinetic decoupling before neutrino decoupling, while panel (b) shows the corresponding results for dark sector decoupling after neutrino decoupling. }
  \label{plot_v}
\end{figure}
Here, $\epsilon$ quantifies the change in the entropy of the SM sector due to the injection of SM particles from dark matter annihilation or bound state decay, and it is defined as $\epsilon=\frac{a^3_i s_i^{SM}}{a^3_f s_f^{SM}}$ . Similarly, $\eta$ describes the corresponding change in the entropy of the dark  sector due to dark matter annihilation or bound state decay into dark radiation, and is given by, $\eta=\frac{a^3_i s_i^{DR}}{a^3_f s_f^{DR}}$.

Figures~\ref{plot_s} and~\ref{plot_v} illustrate the impact of exotic energy injections on the effective number of relativistic species, $\Delta N_{\mathrm{eff}}$, for scalar and vector mediated dark matter interactions, respectively. Following Ref.~\cite{ne,planck}, the Planck $2\sigma$ upper bound on $\Delta N_{\mathrm{eff}}$ is $0.287$, we find that the vector mediated scenario violates this bound for both cases, in which the dark sector kinetically decouples before and after neutrino decoupling. In contrast, for scalar mediated dark matter interactions, the constraint can be satisfied when the dark sector kinetically decouples within the temperature range $0.6 < T_{\chi\text{-dec}} < 1~\mathrm{MeV}$ and $m_\chi>30 MeV$, the decoupling temperature which satisfy this bound is more than $0.6 MeV$ for $m_\chi <30 MeV$. Although same amount of entropy injection into the SM and dark radiation sectors occurs through decays and annihilations in both scenarios, the different  degrees of freedom associated with the scalar and vector mediators lead to distinct contributions to $\Delta N_{\mathrm{eff}}$. Consequently, our model remains viable under the $\Delta N_{\mathrm{eff}}$ constraints only in the scalar mediated dark sector scenario, provided that kinetic decoupling occurs within the allowed temperature range.
\subsection{Determination of kinetic decoupling temperature}

In the previous section, we established that the dark sector remains in thermal equilibrium with the Standard Model  bath. We also found that, for the model to remain consistent with the observational constraints on $\Delta N_{\mathrm{eff}}$, the kinetic decoupling temperature must lie within the range $0.6 < T_{\chi\text{-dec}} < 1~\mathrm{MeV}$. In this section, we determine the kinetic decoupling temperature for the dark sector. We
now determine the temperature at which the dark sector
kinetically decouples from the Standard Model thermal
bath. This decoupling is characterized by the momentum
transfer rate
\begin{equation}
\gamma(T)=
\frac{1}{48\pi^{3} g_{\chi} T m_{\chi}^{3}}
\int d\omega \,
k^{4}\,
f_{k}(1\pm f_{k})\,
\langle |\mathcal{M}|^{2}\rangle ,
\label{eq:gamma}
\end{equation}
where $|\mathcal{M}|$ denotes the $t$-channel matrix
element for DM--SM scattering, and  $f_k$ is the statistical distribution function, $f_k = \left(e^{(E-\mu_c)/T} \pm 1\right)^{-1}$, where the $+$ ($-$) sign corresponds to the Fermi--Dirac (Bose-Einstein) distribution and $\mu_c$ is the chemical potential. The thermally
averaged momentum-transfer matrix element is defined as
\begin{equation}
\langle |\mathcal{M}|^{2}\rangle
=
\frac{1}{8k^{4}}
\int_{-4k^{2}}^{0}
(-t)\,
|\mathcal{M}|^{2}\,
dt,
\label{eq:Msqavg}
\end{equation}
with $t$ being the Mandelstam variable. Defining
$T_{\chi}$ as the dark sector temperature, the kinetic
decoupling temperature is obtained by solving
\begin{equation}
\frac{dT_{\chi}}{dT}
-
2\frac{T_{\chi}}{T}
=
\frac{\gamma(T)}{H(T)}
\left(
\frac{T_{\chi}}{T}-1
\right).
\label{eq:Tchi}
\end{equation}
\begin{figure}[t]
  \includegraphics[width=0.45\textwidth]{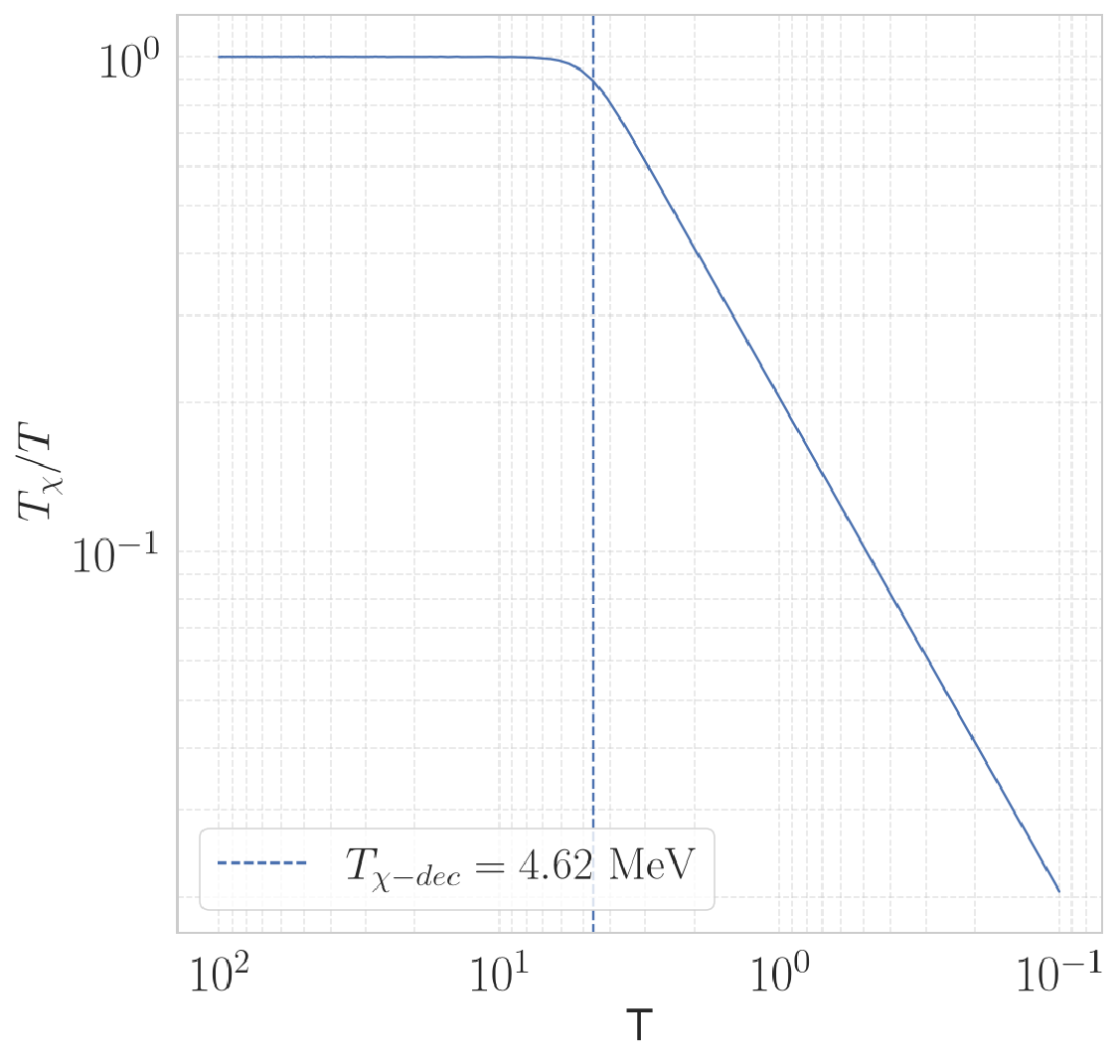}
  \caption{The evolution of the dark sector temperature
$T_\chi$ (in MeV) is shown, with the dotted line indicating
the decoupling temperature.}
\end{figure}
\begin{figure}[!htbp]
  \centering

  \begin{minipage}{0.9\linewidth}
        \centering
        \text{(a)}\\[2mm]
        \includegraphics[width=\linewidth]{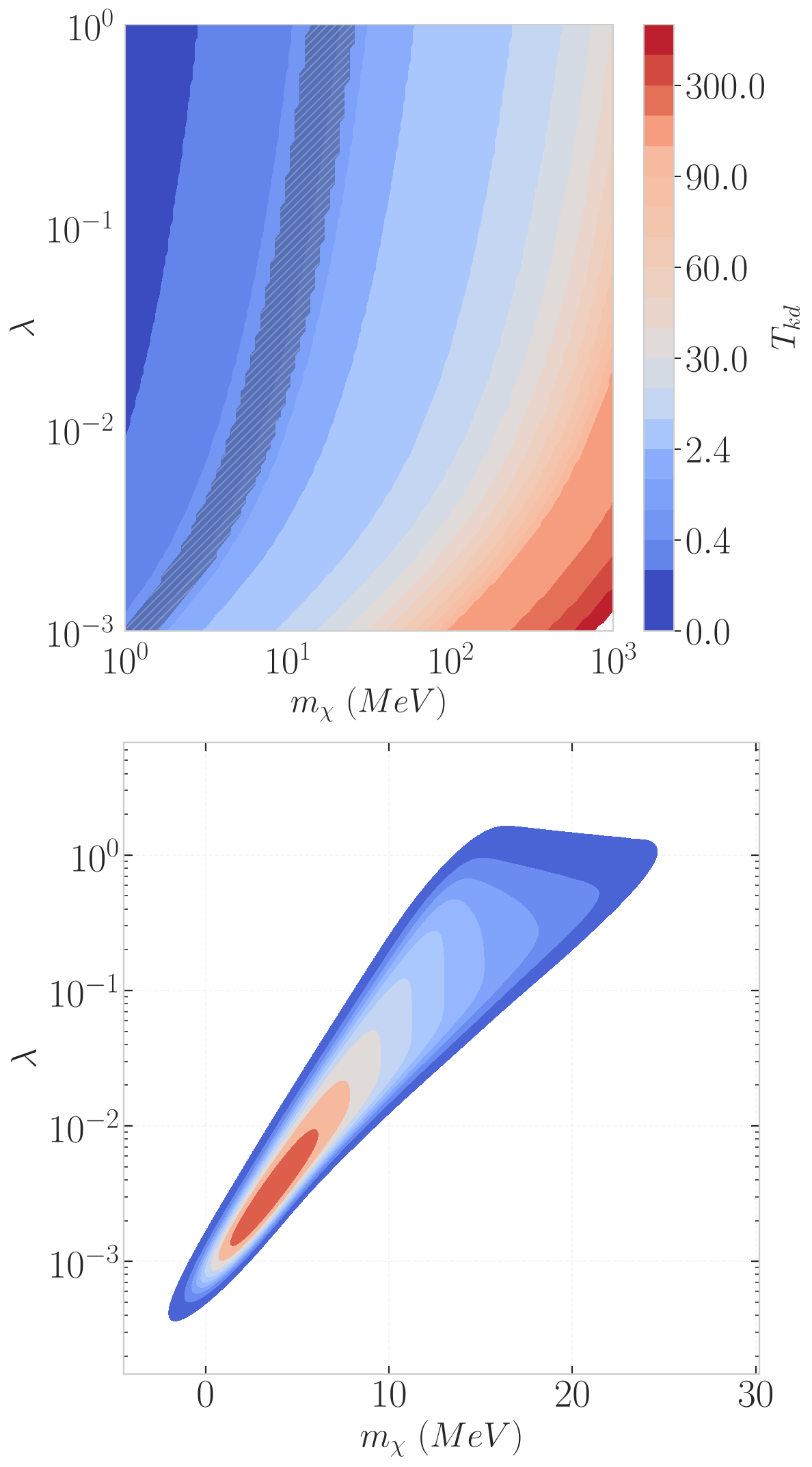}\\[2mm]
        \text{(b)}
    \end{minipage}

  \caption{Panel (a) shows the parameter scan in the dark matter mass and DM-SM coupling strength plane, with the shaded region indicating parameter points that satisfy the $\Delta N_{\mathrm{eff}}$ constraints on the kinetic decoupling temperature. Panel (b) shows the kernel density estimation of the accepted parameter region }
  \label{plot_tkd}
\end{figure}
This equation shows that for $\gamma \gg H$, the dark
sector remains in kinetic equilibrium with the Standard
Model bath, such that $T_{\chi}\simeq T$. As the Universe
expands and the interaction rate drops to
$\gamma\sim H$, $T_{\chi}$ begins to deviate from the bath
temperature, signaling kinetic decoupling of the dark
sector. As shown in Figure ~9, for
$m_{\chi}=150~\mathrm{MeV}$, the dark sector departs from
kinetic equilibrium at a temperature of
$T=4.62~\mathrm{MeV}$, which occurs after freeze-out.
This indicates that the dark sector remains in thermal
equilibrium until freeze-out, and it decouples from the
Standard Model thermal bath shortly thereafter. After
decoupling, the dark sector temperature redshifts as
$T_{\chi}\propto a^{-2}.$

Figure~\ref{plot_tkd} shows the parameter scan in the $\lambda$--$m_\chi$ plane, we observe that larger values of $\lambda$ lead to delayed kinetic decoupling, as stronger scattering maintains kinetic equilibrium for a longer duration. Conversely, the kinetic decoupling temperature increases with increasing dark matter mass. Furthermore, we find that only dark matter candidates within the mass range $1 < m_\chi < 25~\mathrm{MeV}$ satisfy the $\Delta N_{\mathrm{eff}}$ constraints established in the previous section. The KDE analysis further indicates that parameter points with $m_\chi < 10~\mathrm{MeV}$ and $\lambda < 10^{-2}$ are more densely concentrated, suggesting that these parameters are more likely to yield kinetic decoupling temperatures in range $0.6 < T_{\chi\text{-dec}} < 1~\mathrm{MeV}$.

The presence of $\chi$, $A^{\mu}$, $B_n$, and their
interactions with the Standard Model would, in general,
introduce new degrees of freedom and exotic energy
injection into the thermal plasma. These effects could
potentially modify the temperature evolution of the
Universe, alter the standard BBN predictions and leave
imprints on the CMB anisotropy spectrum.  However, a detailed
analysis of the constraints from BBN and CMB, and
 is beyond the scope of the present
work and will be addressed in a forthcoming study.

\section{Indirect detection}
\label{sec:id}
As discussed in the previous section, for Sub-GeV dark matter candidates the only kinematically accessible annihilation channels are 
$\chi\chi \rightarrow e^+ e^-$, $\chi\chi \rightarrow \mu^+ \mu^-$, and $\chi\chi \rightarrow \pi^+ \pi^-$. 
We do not consider annihilation into neutral pion states, since the resulting signals are too faint to be detected by Sub-GeV  telescopes~\cite{IDs1}.
In this section, we compute the photon flux produced from both the annihilation of dark matter and the decay of SIMPonium into visible particles. 
The differential gamma ray flux produced by dark matter annihilation  is given by \cite{IDs2},
\begin{equation}
\frac{d\Phi_{\rm ann}}{dE_\gamma\, d\Omega}
=
\frac{1}{8\pi m_\chi^2}\,
\langle\sigma v\rangle_{\chi\chi \rightarrow f\bar{f}}\,
\frac{dN_\gamma^f}{dE_\gamma}\,
J(\theta),
\label{eq:flux_ann}
\end{equation}
where $m_\chi$ is the dark matter mass, $\langle\sigma v\rangle_{\chi\chi \rightarrow f\bar{f}}$ is the thermal averaged annihilation cross section of dark matter annihilation  into the final state $f\bar f$ and $dN_\gamma^f/dE_\gamma$ denotes the photon spectrum produced per annihilation in that channel.

For decaying dark matter, the corresponding expression is,
\begin{equation}
\frac{d\Phi_{\rm dec}}{dE_\gamma\, d\Omega}
=
\frac{1}{4\pi m_\chi}\,
\Gamma_{B_n \rightarrow f\bar{f}}\,
\frac{dN_\gamma^f}{dE_\gamma}\,
D(\theta),
\label{eq:flux_dec}
\end{equation}
where $\Gamma_{\chi \rightarrow f\bar{f}}$ is the decay rate of the dark matter particle.

The quantities $J(\theta)$ and $D(\theta)$ encode the astrophysical dependence of the signal through line of sight (l.o.s.) integrals over the dark matter density distribution $\rho(r)$,
\begin{equation}
J(\theta) = \int_{\rm l.o.s.} \rho^2\!\big(r(s,\theta)\big)\, ds,
\qquad
D(\theta) = \int_{\rm l.o.s.} \rho\!\big(r(s,\theta)\big)\, ds,
\label{eq:J_D_def}
\end{equation}
where $s$ is the distance along the line of sight, $\rho\!\big(r(s,\theta)\big)$ is the dark matter density  and for the analyses that follow, we use the $J(\theta)$ and $D(\theta)$ factors corresponding to the Milky Way Galactic Centers.
Since only a fraction of the total dark matter abundance is in the form of decaying bound states, the $D$-factor must be modified to account for their fractional contribution. Accordingly, we define the effective $D$-factor as
\[
D'(\theta)=\int_{\rm l.o.s.} f_B\,\rho\!\big(r(s,\theta)\big)\,ds
=f_B D(\theta),
\]
where $f_B=\rho_B/\rho_\chi$ denotes the fraction of dark matter density contributed by the bound states. In this section, we investigate the indirect detection signatures arising from SM particle injection through the decays of ground-state and excited SIMPonium states, as well as the annihilation of free dark matter particles, fixing $f_B=\mathcal{O}(10^{-2})$ and $\lambda=g_\chi=\mathcal{O}(10^{-3})$. 
The  term $\frac{dN_\gamma^f}{dE_\gamma} $ receives several contributions, such as inverse Compton scattering, final-state radiation, 
radiative decay, and bremsstrahlung emission. In this work, we focus on predicting the photon spectra 
arising from final state radiation and radiative decay,
\begin{equation*}
    \frac{dN^f_\gamma}{dE_\gamma}=\frac{dN^f_\gamma}{dE_\gamma} \bigg|_{FSR}+\frac{dN^f_\gamma}{dE_\gamma}\bigg|_{RD}.
\end{equation*}

\subsection{Final-state radiation}
The annihilation of free dark matter or decay of bound states into charged Standard Model particles produces photon via final state radiation  and the resulting spectra due to leptonic final states is given by \cite{fsr},
 \begin{equation}
     \frac{dN^{l^+l^-}}{dE_\gamma }\bigg|_{FSR}=\frac{\alpha}{\pi\beta(3-\beta^2)m_\chi} \bigg[A \ln \frac{1+R(\nu)}{1-R(\nu)}-2R(\nu)B\bigg],
 \end{equation}
 where $l=e,\mu$,
\( A = \frac{(1+\beta^2)(3-\beta^2)}{\nu} - 2(3-\beta^2) + 2\nu \;\), 
\( B = \frac{(1-\nu)(3-\beta^2)}{\nu} + \nu \;\), 
\( \beta = 1 - 4\eta^2 \;\), 
\( \nu = E_\gamma / m_\chi \;\), 
\( \eta = \frac{m_l}{2m_\chi} \;\), and 
\( R(\nu) = \sqrt{1 - \frac{4\eta^2}{1 - \nu}} \;\).
Similarly, FSR spectrum from charged pions is given by,
  \begin{align}
     \frac{dN^{\pi^+\pi^-}}{dE_\gamma }&=\frac{2\alpha}{\pi\beta m_\chi} \bigg[\bigg(\frac{\nu}{\beta^2}-\frac{1-\nu}{\nu}\bigg)R(\nu) \nonumber \\
     &\quad +\bigg(\frac{1+\beta^2}{2\nu}-1\bigg)\ln \frac{1+R(\nu)}{1-R(\nu)}\bigg],
 \end{align}
where $\nu ,\beta$ and  $\eta$ are defined as above with $m_l$ replaced by $m_\pi$.

\begin{figure}[t]
  \centering
  \includegraphics[width=0.8\linewidth]{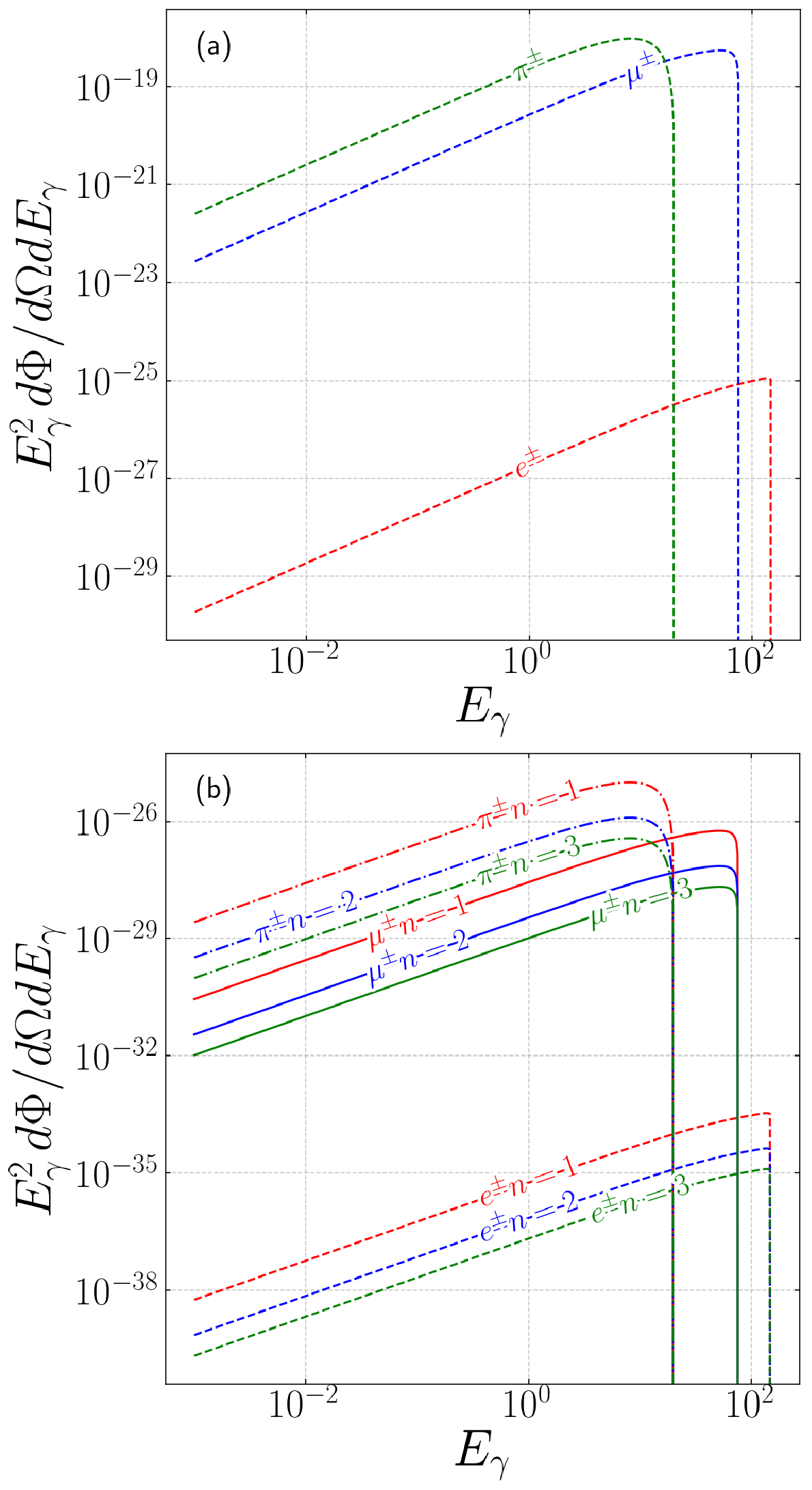}
  \caption{Panel (a) shows the final state radiation spectrum produced by charged leptons and pion which was product of annihilation of free $\chi $ as a function of energy of emitted photon in MeV. 
  Panel (b) illustrates the FSR that was produced due to decay of SIMPonium into SM particles, where the red, green, and blue curves correspond to decay of $n=1$, $n=2$, and $n=3$ states, respectively. Solid, dashed, and dotted lines represent the final SM particles as  $\mu^{\pm}$,$e^{\pm}$ and $\pi^{\pm}$, respectively.}
  \label{plot_fsr}
\end{figure}
In Fig.~\ref{plot_fsr}(a), we show the differential flux of final state radiation produced by $e^\pm$, $\mu^\pm$, and $\pi^\pm$. We consider a dark matter mass of $m_\chi = 150~\mathrm{MeV}$ and the differential flux produced by it exhibits a sharp cutoff at 
$E_\gamma^{\mathrm{max}} = m_\chi \left( 1 - \frac{m_f^2}{m_\chi^2} \right)$, 
where $m_f = m_e,\, m_\mu,\, m_\pi$. Accordingly, the cutoff energies for $e^\pm$, $\mu^\pm$, and $\pi^\pm$ are approximately $E_\gamma^{\mathrm{max}} \simeq 149~\mathrm{MeV}$, $75~\mathrm{MeV}$, and $20~\mathrm{MeV}$, respectively. Also  it is clear that X-ray and soft gamma ray  produced from pions and muons dominate the overall flux, while muon produce slightly more energetic photons. Even if the photon flux produced by electrons are very small, they leave a distinctive feature by producing the most energetic gamma rays among all final states.

The FSR spectra arising from the decay of SIMPonium are shown in Fig.~\ref{plot_fsr}(b). We observe that the differential flux corresponding to visible decays are generally smaller than those produced by the annihilation of free $\chi$ particles. The photon energy cutoff follows a similar pattern as in the  annihilation case, with $E_\gamma^{\mathrm{max}} = 149~\mathrm{MeV},\, 75~\mathrm{MeV},\, 20~\mathrm{MeV}$ for the three decay channels. Since the decay rate satisfies $\Gamma \propto |\psi_n(r=0)|^2$, the ground states decay more efficiently than the excited states. Similar to the flux produced due to  annihilation  the photon flux from bound state decay is also dominated by the FSR contributions from muons and pions.

While the results  correspond to $m_\chi = 150~\mathrm{MeV}$, for $m_\chi < 110~\mathrm{MeV}$ the photon flux becomes very small, as only FSR from $e^\pm$ is kinematically allowed, and this channel produces a comparatively weaker photon flux. 
\begin{figure}[t]
  \centering
  \includegraphics[width=0.85\linewidth]{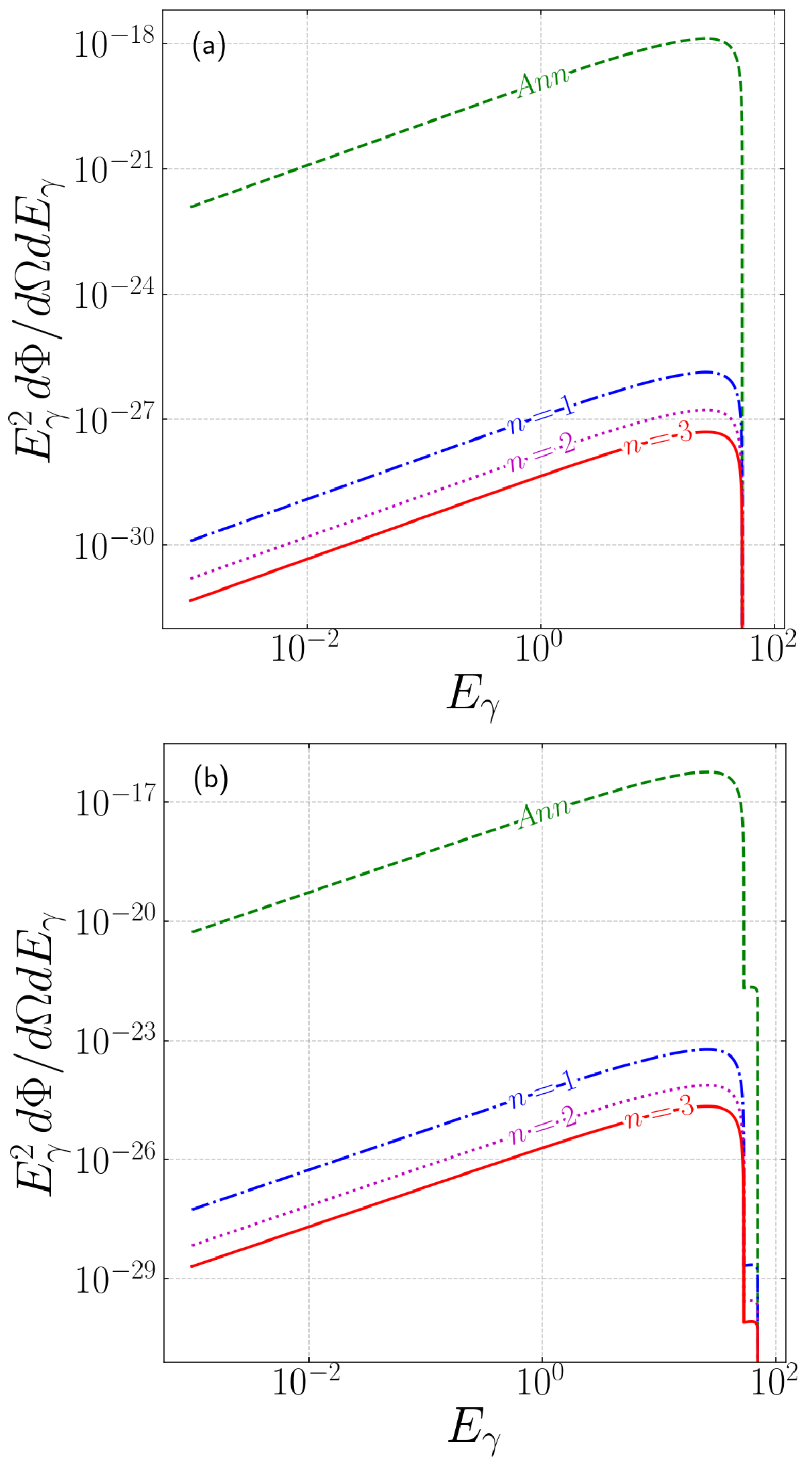}
  \caption{Panel (a) shows the radiative decay spectrum produced by decay of unstable
muon as a function of energy of emiited photon in MeV. 
  Panel (b) illustrates the total radiative decay spectrum
combining the contribution from both muon and pion decay. In both panels  the green, blue, purple and red curves correspond to annihilation of free dark matter and  decay of $n=1$, $n=2$, and $n=3$ states, respectively. }
  \label{plot_rad_muon}
\end{figure}
\subsection{Radiative decay}
When dark matter candidates annihilate or decay into an unstable Standard Model particle, the latter can subsequently produce photons through a radiative decay of the form $\mathrm{SM} \rightarrow \mathrm{SM} + \gamma$. For example, in the radiative decay of the muon,
$\mu^- \rightarrow e^-\, \bar{\nu}_e\, \nu_\mu\, \gamma$, the resulting photon spectrum in muon's rest frame  is given by \cite{fsr2},
\begin{equation}
        \begin{gathered}
    \frac{dN^\mu_{rad}}{dE_\gamma }=\frac{\alpha(1-x)}{36\pi E_\gamma }\bigg[12(3-2x(1-x)^2)\log \frac{1-x}{r} \\
    +x(1-x)(46-55x)-102\bigg], 
        \end{gathered}    
 \end{equation}
 where, $x=\frac{2E_\gamma}{m_\mu}$ and $r=(\frac{m_e}{m_\mu})^2$.

The differential flux produced by radiative decay of muons is shown in Fig.~\ref{plot_rad_muon}(a) and  differential flux   arising from the annihilation of free $\chi$ particles dominates over the flux  produced by the decay of SIMPonium and the spectrum cutoff sharply at $E_{\gamma}^{max}\approx 75$ MeV.

\bigskip
\begin{figure}[t]
  \centering
  \includegraphics[width=0.85\linewidth]{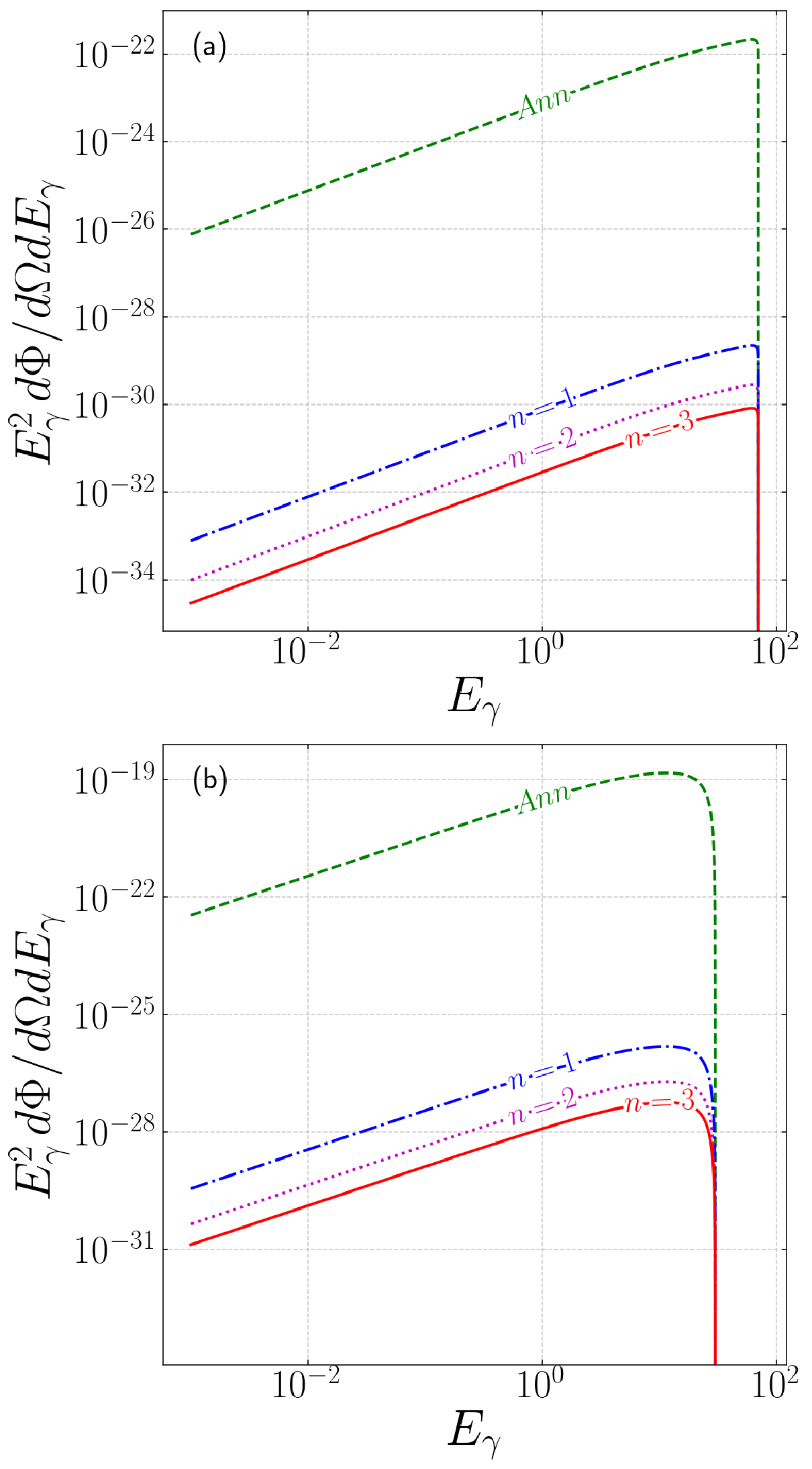}
   \caption{Panel (a) shows the radiative decay spectrum produced by unstable pions decaying to electrons as a function of energy of emitted photon in MeV. 
  Panel (b) illustrates radiative decay spectrum produced by  unstable pions decaying to muons. In both panels  the green, blue, purple and red curves correspond to annihilation of free dark matter and  decay of $n=1$, $n=2$, and $n=3$ states, respectively. }
  \label{plot_rad_pion}
\end{figure}
 Similarly, charged pions can produce photons through the radiative decay channel 
$\pi^- \rightarrow \ell^-\, \bar{\nu}_\ell\, \gamma$, and the corresponding photon spectrum is given by,
\begin{equation}
     \frac{dN^\pi_{rad}}{dE_\gamma }=\frac{\alpha[f(x)+g(x)]}{24\pi m_\pi f^2_\pi(r-1)^2(x-1)^2 r x  },
 \end{equation}
 where,

\begin{align}
f(x) &= (r + x + 1)\Bigl[
m_\pi^2 x^4 (F_A^2 + F_V^2)
\left(r^2 - rx + r \right. \nonumber\\
&\qquad \left. - 2(x - 1)^2\right)
- 12\sqrt{2}\, f_\pi m_\pi r (x - 1) x^2 \nonumber\\
&\qquad \left(F_A(r - 2x + 1) + x F_V\right)
- 24 f_\pi^2 r (x - 1) \nonumber \\
&\qquad \left(4r(x - 1) + (x - 2)^2\right) \Bigr],
\end{align}

  \begin{align}
    g(x) &= 12\sqrt{2} f_\pi r (x - 1)^2 \log\left(\frac{r}{1 - x}\right)\Bigl[m_\pi x^2 \left(F_A(x - 2r) \right. \nonumber \\
    &\qquad \left. - x F_V\right) + \sqrt{2} f_\pi (2r^2 - 2rx) - x^2 + 2x - 2\Bigr],
  \end{align}

 where $x=\frac{2E_\gamma}{m_\pi}$, $r=(\frac{m_l}{m_\pi})^2$ and $f_\pi$is pion decay constant, its value is 92.2 MeV. The axial and vector form factors are given by $F_A=0.0119$ and $F_V(q^2)=F_V(0)(1+aq^2) $ with $F_V(0)=0.0254,a=0.10,q=(1-x)$.

 \bigskip                                                                                                                                                             
If the pion first decays into an on-shell muon, the resulting muon subsequently undergoes radiative decay and produces photons. Taking this possibility into account, the total photon spectrum arising from pion decay is,
\begin{align}
\left.\frac{dN_{{\rm RadTot}\, \gamma}^{\pi}}{dE_\gamma}\right|_{E_\pi = m_\pi} &=
\sum_{\ell = e,\, \mu} {\rm BR}\!\left(\pi \rightarrow \ell \nu_{\ell}\right)
\left.\frac{dN_{{\rm Rad}\, \gamma}^{\pi}}{dE_\gamma}\right|_{E_\pi = m_\pi} \nonumber \\
&\quad + {\rm BR}\!\left(\pi \rightarrow \mu \nu_{\mu}\right)
\left.\frac{dN_{{\rm Rad}\, \gamma}^{\mu}}{dE_\gamma}\right|_{E_\mu = E_\star},
\label{eq:pion_radtot}
\end{align}
where $E_{\pi}$ and $E_{*}$ represent the pion and muon energies in the pion rest frame.

The two separate contributions $(l = e, \mu)$ corresponding to the first term in Eq.~\eqref{eq:pion_radtot} are shown in Figs.~\ref{plot_rad_pion}(a) and (b) and since ${\rm BR}\!\left(\pi \rightarrow \mu \nu_{\mu}\right) \approx 0.98$  the second term is identical to the radiative decay from muons as shown in Fig.~\ref{plot_rad_muon}(a). And due to this high branching ratio, photon production during pion decay into muons dominates over the electron channel. The maximum photon energy during radiative decay is determined by 
$E_\gamma^{\mathrm{max}} = \frac{m_\pi^2 - m_f^2}{2 m_\pi}$, 
where $f = e, \mu$. Using this, we estimate  the maximum photon energies for pion decays into muons and electrons as $E_\gamma^{\mathrm{max}} \simeq 74$ and $35~\mathrm{MeV}$, respectively.  

Fig.~\ref{plot_rad_muon} (b) shows the complete radiative decay spectrum obtained by combining the contributions from both muon decay and pion decay. It is evident that the differential flux is dominated by muon decay and the muonic decay channel of the pion. It is also notable that the photon flux produced by radiative decay is greater than that produced by final-state radiation. Similar to the FSR spectrum, the radiative decay spectrum decreases drastically for $m_\chi < 100~\mathrm{MeV}$ due to kinematic suppression.
\subsection{Constraints from observational data}
\begin{figure}[t]
  \centering
  \includegraphics[width=\linewidth]{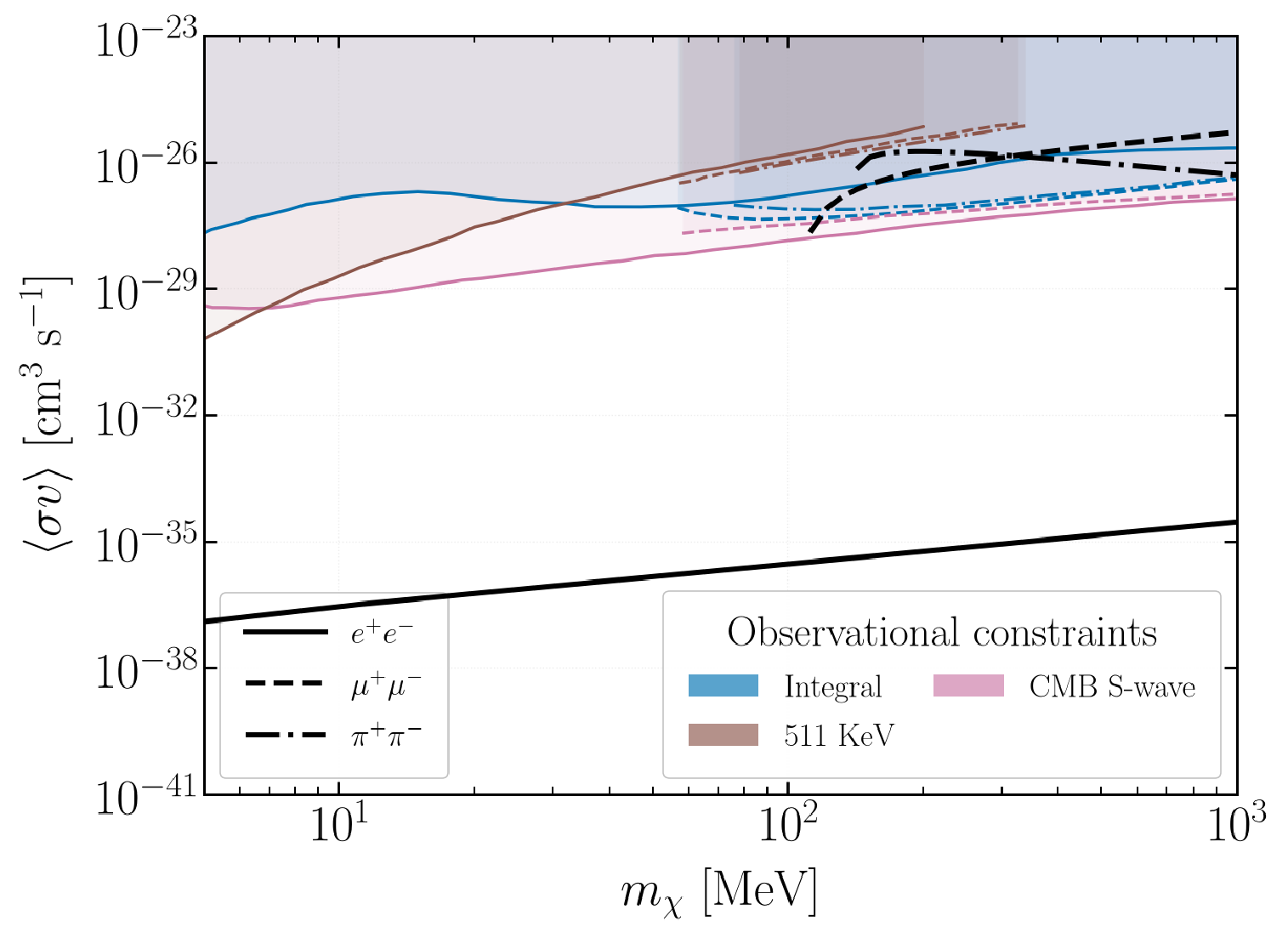}
   \caption{Constraints on the thermally averaged dark matter annihilation cross section derived from CMB observations, the 511~keV line, and Integral X-ray Telescope data. }
  \label{conx}
\end{figure}
\begin{figure*}[t]
    \centering
    \includegraphics[width=\textwidth]{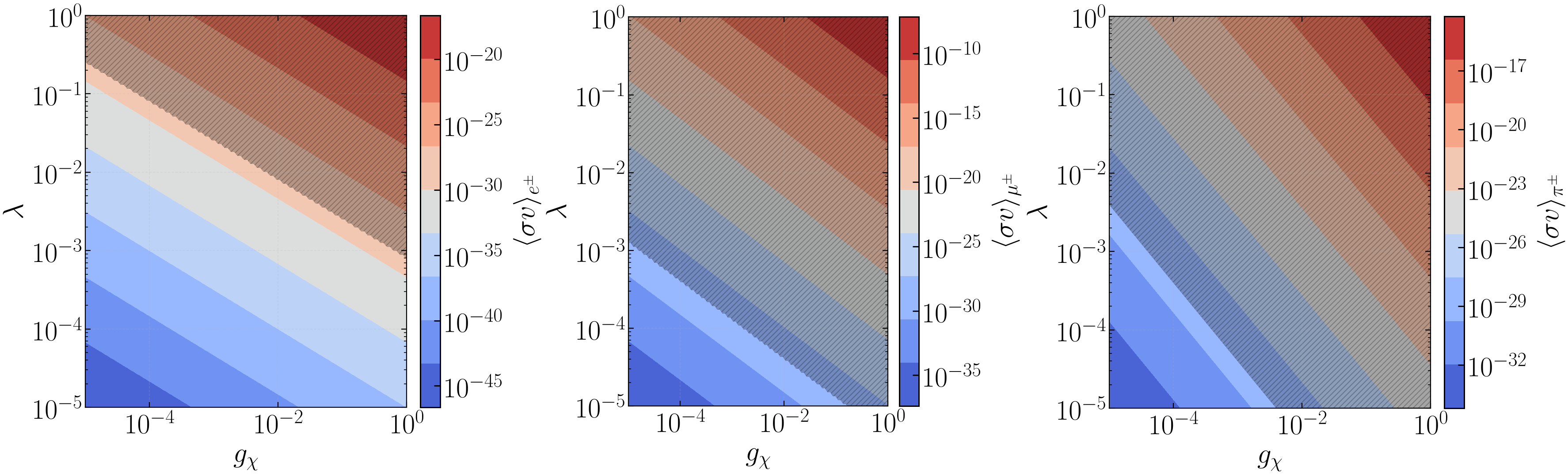}
    \caption{$\lambda - g_\chi$ parameter space constraints for annihilation into kinematically allowed Standard Model particles for a dark matter candidate with mass $m_\chi = 150\text{ MeV}$}
        \label{plot_co}
\end{figure*}

In the previous section, we calculated the differential flux for a fixed set of benchmark parameters. In this section, we explore the parameter space spanned by $g_\chi$ and $\lambda$ to evaluate the thermally averaged annihilation cross section into Standard Model  particles and further constrain the parameter space using observational data. Specifically, we consider constraints from cosmic microwave background  observations of $s$-wave annihilation \cite{c1,c2}, the 511~keV line \cite{ke}, and Integral X-ray Telescope observations \cite{int2}. Before exploring the parameter space, we first present the constraints imposed by these observations over the dark matter mass range $1~\mathrm{MeV}<m_\chi<1000~\mathrm{MeV}$. We calculate the corresponding annihilation cross sections using the same benchmark couplings adopted in the previous section. The resulting constraints and cross sections are presented in Fig.~\ref{conx}. 
We observe that annihilation into the $e^\pm$ channel is highly suppressed  and this is due to the choice of the Higgs portal as mediator between dark and standard model sector. The Higgs portal couples to SM fermions with a coupling strength proportional to $\frac{m_f}{v_h}$, where $m_f$ is the fermion mass and $v_h$ is the Higgs vacuum expectation value. Since the ratio $\frac{m_f}{v_h}$ is extremely small, the annihilation rate into the electron positron channel is negligible. Consequently for $g_\chi=\lambda=10^{-3}$, the annihilation channels into other kinematically accessible SM final states are excluded by the combined constraints from experimental data, but the $e^\pm$ channel remains unconstrained over the entire dark matter mass range considered.

In Fig.~\ref{plot_co}, we present the parameter-space scan over $g_\chi$ and $\lambda$, with the shaded regions indicating parameter points excluded by the combined constraints from experimental data. The scan is performed for a fixed dark matter mass of $m_\chi=150~\mathrm{MeV}$. As expected, owing to the strong suppression of annihilation into the $e^\pm$ channel, a larger fraction of the parameter space remains compatible with experimental constraints compared with the corresponding annihilation channels into $\mu^\pm$ and $\pi^\pm$. Furthermore, in Figure 16, the KDE distributions of the allowed parameter space for $1~\mathrm{MeV}<m_\chi<1000~\mathrm{MeV}$ indicate that a broadly similar range of $\lambda$ and $g_\chi$ remains consistent with experimental constraints across the mass range considered. Moreover, for $m_\chi<100~\mathrm{MeV}$ only a small fraction of the sampled parameter space is excluded. The constraints from $\Delta N_{\mathrm{eff}}$ and  $T_{\chi\text{-dec}}$, established in the previous sections, already restrict the viable dark matter mass range to $m_\chi<25~\mathrm{MeV}$. Consequently, heavier dark matter candidates are excluded within the framework and parameter ranges considered. In this allowed low mass regime, annihilation of Higgs portal mediated SIMPonium into SM particles is restricted to the $e^\pm$ channel, as the other channels are kinematically inaccessible. The coupling $\lambda$ is constrained by the combined requirements from $\Delta N_{\mathrm{eff}}$, $T_{\chi\text{-dec}}$, and indirect detection observational data, whereas $g_\chi$ remains largely unconstrained. This leaves considerable freedom in the dark matter self-interaction strength within the Higgs portal mediated SIMPonium scenario.
\begin{figure*}[t]
    \centering
    \includegraphics[width=\textwidth]{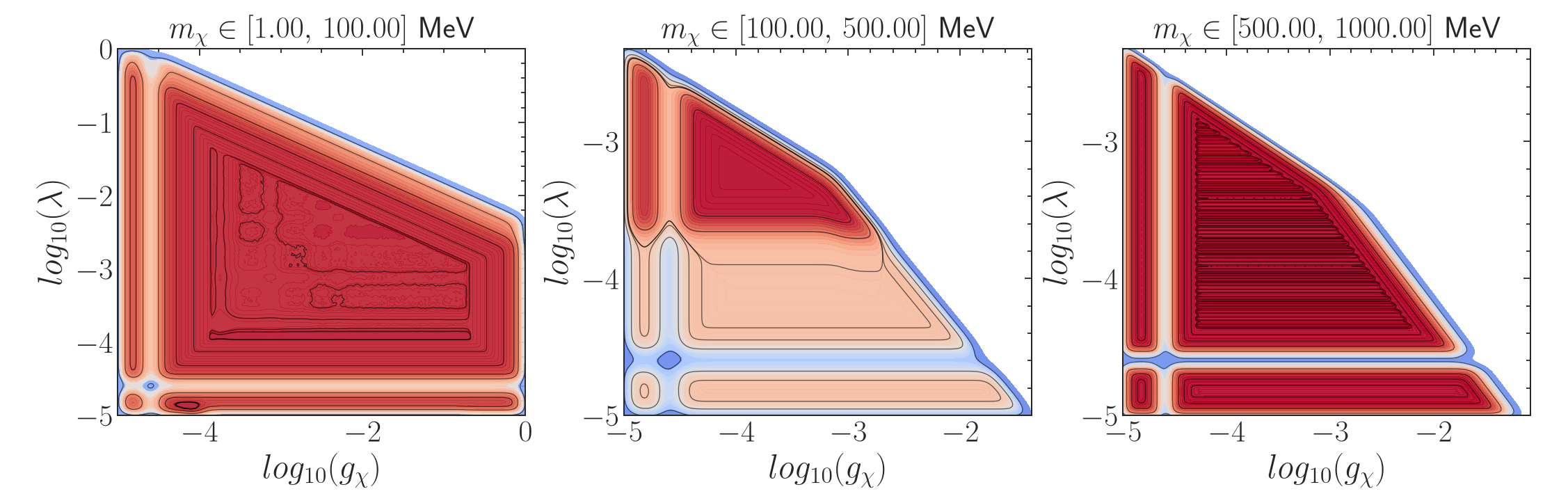}
    \caption{KDE distributions of accepted region in $\lambda - g_\chi$ parameter space}
    \label{plot_cokde}
\end{figure*}
\section{Conclusion}
\label{sec:con}
In this work, we demonstrate that the presence of bound states significantly influences the evolution of the dark matter relic abundance. In particular, the final relic density is determined by the coupled evolution of free dark matter and SIMPonium, with bound-state formation and decay modifying the conventional $4\chi\rightarrow 2\chi$ number-changing process. By employing Monte Carlo simulations and kernel density estimation, we investigate the parameter space and identify regions that successfully reproduce the observed dark matter relic abundance. Our analysis indicates that yields of approximately $Y_\chi\sim 10^{-9}$ for free dark matter and $Y_B\sim 10^{-13}$ for SIMPonium can account for the observed relic density. We also obtain approximate numerical estimates of the thermally averaged cross sections, $\langle \sigma v^n\rangle_i$, associated with the relevant interaction processes, providing insight into the conditions required for a successful freeze out mechanism. Among the processes considered, $\langle \sigma v\rangle_J$ and $\Gamma_I$ exhibit the strongest negative correlations with the final relic abundance, highlighting their significant influence on the dark matter evolution. The free dark matter component undergoes freeze out at $x_f\approx 20$, whereas the SIMPonium abundance does not exhibit a distinct freeze out. Consequently, the present day SIMPonium abundance is significantly smaller than that of free dark matter.

The presence of an additional massless mediator, $\Phi$, contributes to the effective number of relativistic species, $N_{\rm eff}$, through $\Delta N_{\rm eff}$. We find that the observational constraint on $N_{\rm eff}$ can be satisfied only if $\Phi$ is a scalar boson. In contrast, a massless vector mediator yields $\Delta N_{\rm eff}>0.287$, irrespective of whether the dark sector kinetically decouples from the Standard Model plasma before or after neutrino decoupling. To further constrain the model, we estimate the kinetic decoupling temperature of the dark sector and investigate the parameter space consistent with the observational bounds on $N_{\rm eff}$. Our results show that, for a scalar mediator, the constraint on $\Delta N_{\rm eff}$ is satisfied when kinetic decoupling occurs at temperatures between $0.6$ and $1\,\mathrm{MeV}$. Furthermore, we find that this constraint can be satisfied only for dark matter masses $m_\chi<25\,\mathrm{MeV}$.

The indirect detection study shows that lighter dark matter candidates ($m_\chi < 110~\mathrm{MeV}$) produce a much weaker photon spectrum compared to heavier dark matter. The annihilation of free dark matter generates a higher photon flux than the decay of SIMPonium into Standard Model particles, and the photon flux from excited state decays is even smaller.Furthermore, since the Higgs-mediated scenario allows only dark matter particles with masses $m_\chi<25\,\mathrm{MeV}$, the resulting photon flux is too small to be constrained by existing observations, including the $511\,\mathrm{keV}$ line, cosmic microwave background  constraints on $s$-wave annihilation, and measurements from the INTEGRAL X-ray telescope. This weak photon signal primarily arises from the suppressed coupling between dark matter and electrons, which significantly reduces the corresponding annihilation and decay rates into $e^\pm$.
\onecolumngrid

\appendix
\section{Coulomb wavefuctions}
\label{app:A}
\subsection{Bound state wavefucntion}
For a two body system interacting via an attractive Coulomb potential \(V(r) = -V_0/r\), the radial Schrödinger equation describing the relative motion in the center of mass frame (with \(\hbar = 1\)) takes the form given by Eq.~\ref{eq:sc}. We first derive the solutions corresponding to the  bound states of this system,

\begin{equation}
  \label{eq:sc}
    \frac{d^2 R(r)}{dr^2} + \frac{2}{r} \frac{dR(r)}{dr}+2 \mu[E+\frac{V_0}{r}-\frac{l(l+1)}{2\mu r^2}]R(r)=0.
  \end{equation}
Substituting $r=s$ and $R(r)=\psi(s)$ the above equation simplifies as,
\begin{equation}
    \frac{d^2 \psi(s)}{ds^2} + \frac{2}{s} \frac{d\psi(s)}{ds}+\frac{1}{s^2}[-\alpha s^2+\beta s - \gamma]=0,
\end{equation}
where,$\alpha=-2\mu E$, $\beta=2\mu V_0$ and $\gamma=l(l+1)$.

Following the Nikiforov--Uvarov method to solve this differential equation, we obtain the radial wavefunction in the form,
\begin{equation}
    R(r)=N_{nlm} r^l e^{-\sqrt{\alpha}r} L_n^{2l+1}(2r\sqrt{\alpha}),
\end{equation}
where $ L_n^{2l+1}(2r\sqrt{\alpha})$ is the associated laguerre polynomial and the complete wave function is given by,
\begin{equation}
    \psi_n(r)=R(r) Y^l_m,
\end{equation}
and the energy eigenvalues are given by,
\begin{equation}
    E_n=\frac{-V_0^2 \mu}{2 n^2}.
\end{equation}
\subsection{Scattering state wavefucntion}
To describe the scattering states, we use the relation,
\begin{equation*}
    E = \frac{P^2}{2\mu} = \frac{\mu v^2}{2},
\end{equation*}
and the Schrödinger equation for the Coulomb interaction can be written as,
\begin{equation}
\left[\nabla^2 + P^2 - \frac{2\delta P}{r}\right]\psi(\mathbf{r}) = 0,
\end{equation} 

where \(\delta = \frac{V_0}{v}\) and \(P = \mu v\). Using the ansatz \(\psi(\mathbf{r}) = e^{iPz} f(r - z)\) and defining the variable \(u = r - z\), the equation simplifies to,
\begin{equation}
    u\frac{d^2f(u)}{du^2}+(1-iPu)\frac{df(u)}{du}+\delta P f(u)=0. 
\end{equation}
This equation is of the confluent hypergeometric type, and its solution can therefore be written in the form,
\begin{equation}
    \psi_s(r)=N e^{iPr} {}_1F_1[-i\delta;1;i(Pr-P.r)].
\end{equation}

\section{Calculation of thermal average cross sections}

\label{app:B}
\subsection{$4\chi \rightarrow 2\chi$ annihilation cross sections }
We derive the thermally averaged cross section for the process \(4\chi \rightarrow 2\chi\). Consider the process,
\begin{equation*}
    \chi^*(p_1) + \chi^*(p_2) + \chi(p_3) + \chi(p_4) \rightarrow \chi(p_5) + \chi(p_6) \, .
\end{equation*}
The thermal average cross section for this process would be given by,
\begin{align}
\langle \sigma v^3 \rangle_{C} &=
\frac{1}{n_1^{\text{eq}} n_2^{\text{eq}} n_3^{\text{eq}} n_4^{\text{eq}}}
\int \frac{g_\chi\, d^3p_1}{(2\pi)^3 2E_1}
     \frac{g_\chi\, d^3p_2}{(2\pi)^3 2E_2}
     \frac{g_\chi\, d^3p_3}{(2\pi)^3 2E_3}
     \frac{g_\chi\, d^3p_4}{(2\pi)^3 2E_4}
     \frac{g_\chi\, d^3p_5}{(2\pi)^3 2E_5}
     \frac{g_\chi\, d^3p_6}{(2\pi)^3 2E_6}\nonumber \\ 
&\quad \times (2\pi)^4 \delta^4(p_1 + p_2 +p_3 + p_4-p_5-p_6)\,
| \mathcal{M} |^2\,
f_1^{\text{eq}} f_2^{\text{eq}} f_3^{\text{eq}} f_4^{\text{eq}} , \nonumber \\ 
&= \frac{1}{n_1^{\text{eq}} n_2^{\text{eq}} n_3^{\text{eq}} n_4^{\text{eq}}}
\int \frac{g_\chi\, d^3p_1}{(2\pi)^3 2E_1}
     \frac{g_\chi\, d^3p_2}{(2\pi)^3 2E_2}
     \frac{g_\chi\, d^3p_3}{(2\pi)^3 2E_3}
     \frac{g_\chi\, d^3p_4}{(2\pi)^3 2E_4}
     (\sigma v^3)_C f_1^{\text{eq}} f_2^{\text{eq}} f_3^{\text{eq}} f_4^{\text{eq}} .
\label{thcs_free1} 
    \end{align}
We know that the number density at equilibrium  would be given by,
\begin{equation*}
  n^{eq}_i=\frac{g_i}{(2\pi)^3}\int d^3p f^{eq}_i.
\end{equation*}
Since dark matter is a cold species $f^{eq}_i=e^{\frac{-E_i}{T_i}}$ and the number density will be,

\begin{equation*}
  n_i^{eq}=\frac{g_i}{(2\pi)^3}4\pi m_i^2 T K_2(m_i/T) ,
\end{equation*}
where $K_2(m_i/T)$ is the Bessel function of second kind. Substituting these in  Eq.~\eqref{thcs_free1} 
we get,
\begin{equation}
  \langle \sigma v^3 \rangle_{C} =\left( \frac{K_1(m_\chi/T)}{K_2(m_\chi/T)} \right)^{\!4}
  (\sigma v^3)_{C}.
  \label{thcs_C}
\end{equation}
Under non relativistic approximation, $x=\frac{m}{T}\gg 1$ \cite{ap},
\begin{equation*}
  \frac{K_1^4(x)}{K_2^4(x)} \approx 1 + \mathcal{O}(x^{-2}),
\end{equation*}
and the thermal average cross section simplifies to,
\begin{equation}
  \langle \sigma v^3 \rangle_{C} \approx (\sigma v^3)_{C}.
  \label{thcs_C2}
\end{equation}
The SIMPonium bound state in the final state is formed via the process,
\begin{equation*}
\chi^*(p_1) + \chi^*(p_2) + \chi(p_3) + \chi(p_4) 
\;\longrightarrow\; 
B_n(p_5) + A^{\mu}(p_6) \, .
\end{equation*}
The thermally averaged cross section is then given by,
\begin{align*}
\langle \sigma v^3 \rangle_{D} 
&= 
\frac{1}{n_1^{\text{eq}} n_2^{\text{eq}} n_3^{\text{eq}} n_4^{\text{eq}}}
\sqrt{\frac{2}{\mu}}
\int 
\frac{g_\chi\, d^3p_1}{(2\pi)^3 2E_1}
\frac{g_\chi\, d^3p_2}{(2\pi)^3 2E_2}
\frac{g_\chi\, d^3p_3}{(2\pi)^3 2E_3}
\frac{g_\chi\, d^3p_4}{(2\pi)^3 2E_4}
\frac{g_{B_n}\, d^3p_5}{(2\pi)^3 2E_5}
\frac{g_{\Phi}\, d^3p_6}{(2\pi)^3 2E_6} \notag \\
&\quad \times 
\frac{d^3Q_1}{(2\pi)^3}
\frac{d^3Q_2}{(2\pi)^3}
(2\pi)^4 \delta^4(p_1 + p_2 +p_3 + p_4-p_5-p_6)\,
|\psi_s(Q_1)|^2\, |\psi_n(Q_2)|^2\,
|\mathcal{M}|^2\,
f_1^{\text{eq}} f_2^{\text{eq}} f_3^{\text{eq}} f_4^{\text{eq}} , \notag \\[1em] 
&= 
\frac{1}{n_1^{\text{eq}} n_2^{\text{eq}} n_3^{\text{eq}} n_4^{\text{eq}}}
\int 
\frac{g_\chi\, d^3p_1}{(2\pi)^3 2E_1}
\frac{g_\chi\, d^3p_2}{(2\pi)^3 2E_2}
\frac{g_\chi\, d^3p_3}{(2\pi)^3 2E_3}
\frac{g_\chi\, d^3p_4}{(2\pi)^3 2E_4}
(\sigma v^3)_D\,
f_1^{\text{eq}} f_2^{\text{eq}} f_3^{\text{eq}} f_4^{\text{eq}} ,
\label{thcs_free}
\end{align*}
\begin{equation}
  \label{thcs_D}
  \langle \sigma v^3 \rangle_{D} =\left( \frac{K_1(m_\chi/T)}{K_2(m_\chi/T)} \right)^{\!4}
  (\sigma v^3)_{D}
   ~ \approx (\sigma v^3)_{D}.
\end{equation}

In Eq.~\eqref{thcs_D}, we did not account for the Bose enhancement 
factor $(1 + f_A(\omega))$, which arises due to the emission of the dark photon 
with energy $\omega = E_n + \frac{\mu v_{\rm rel}^2}{2}$. Let $z = \frac{E_n}{T}$ 
and $\beta = \frac{V_0}{v_{\rm rel}}$, and the  enhancement factor for 
the bound state formation process is given by,
\begin{equation}
    S^n_{\rm BSF}(\beta) = \frac{1}{n} 
    \frac{2^{10}\, \pi \left(\beta/n\right)^5}
    {3\left(1 + \left(\beta/n\right)^2\right)} 
    \frac{e^{-4\beta\,\text{arccot}(\beta/n)}}{1 - e^{-2\pi(\beta/n)}}
    \label{eq:SBSF}
\end{equation}
Assuming $z = z_\chi=\frac{E_n}{T_\chi}$, the thermal average cross section for process D 
with Bose enhancement is,
\begin{equation}
    \langle \sigma v^3 (1 + f_A(\omega))\rangle_{D} 
    = (\sigma v^3)_{D}\, \bar{S}_{\rm BSF}(z),
    \label{eq:thcs_D_bose}
\end{equation}
where $\bar{S}_{\rm BSF}(z)$ is given by,
\begin{equation}
    \bar{S}_{\rm BSF}(z) = \frac{2}{\sqrt{\pi}} 
    \int_{0}^{\infty} du\; S_{\rm BSF}^n\!\left(\sqrt{\frac{z}{u}}\right) 
    \frac{\sqrt{u}\, e^{z}}{e^{z+u}-1}.
    \label{eq:Sbar}
\end{equation}
It is also possible that a previously formed bound state interacts via a four-point vertex and contributes to the total $(4\chi \rightarrow 2\chi)$ annihilation rate. One such process is,
\begin{equation*}
  \chi^*(p_1) + \chi^*(p_2)+B_n(p_3)  \rightarrow \chi^*(p_4) + \chi(p_5).
\end{equation*}
The corresponding thermal average cross section would be,  
\begin{align*}
\langle \sigma v^2 \rangle_{E} 
&= 
\frac{1}{n_1^{\text{eq}} n_2^{\text{eq}} n_3^{\text{eq}} }
\sqrt{\frac{2}{\mu}}
\int 
\frac{g_\chi\, d^3p_1}{(2\pi)^3 2E_1}
\frac{g_\chi\, d^3p_2}{(2\pi)^3 2E_2}
\frac{g_{B_n}\, d^3p_3}{(2\pi)^3 2E_3}
\frac{g_\chi\, d^3p_4}{(2\pi)^3 2E_4}
\frac{g_\chi\, d^3p_5}{(2\pi)^3 2E_5} \\
&\quad \times 
\frac{d^3Q_1}{(2\pi)^3}
(2\pi)^4 \delta^4(p_1 + p_2 +p_3 - p_4-p_5)\,
|\psi_n(Q_1)|^2\,
|\mathcal{M}|^2\,
f_1^{\text{eq}} f_2^{\text{eq}} f_3^{\text{eq}}, \\[1em]
&= 
\frac{1}{n_1^{\text{eq}} n_2^{\text{eq}} n_3^{\text{eq}} }
\int 
\frac{g_\chi\, d^3p_1}{(2\pi)^3 2E_1}
\frac{g_\chi\, d^3p_2}{(2\pi)^3 2E_2}
\frac{g_{B_n}\, d^3p_3}{(2\pi)^3 2E_3}
(\sigma v^2)_E\,
f_1^{\text{eq}} f_2^{\text{eq}} f_3^{\text{eq}} ,
\end{align*}

\begin{equation}
  \langle \sigma v^2 \rangle_{E} = \left( \frac{K_1(m_\chi/T)}{K_2(m_\chi/T)} \right)^{\!2} \frac{K_1(M_{B_n}/T)}{K_2(M_{B_n}/T)}
  (\sigma v^2)_{E}~ \approx (\sigma v^2)_{E}.
\end{equation}
Furthermore, two incoming SIMPonium bound states can annihilate into free dark matter particles, as shown below,
\begin{equation*}
  B_n(p_1) B_n(p_2) \rightarrow   \chi(p_3) \chi(p_4).
\end{equation*}
The thermal average cross section would be,
\begin{align*}
\langle \sigma v \rangle_{F} 
&= 
\frac{1}{n_1^{\text{eq}} n_2^{\text{eq}}}
\frac{2}{\mu}
\int 
\frac{g_{B_n}\, d^3p_1}{(2\pi)^3 2E_1}
\frac{g_{B_n}\, d^3p_2}{(2\pi)^3 2E_2}
\frac{g_\chi\, d^3p_3}{(2\pi)^3 2E_3}
\frac{g_\chi\, d^3p_4}{(2\pi)^3 2E_4} \\[4pt]
&\quad \times 
\frac{d^3Q_1}{(2\pi)^3}
\frac{d^3Q_2}{(2\pi)^3}
(2\pi)^4 
\delta^4\!\big(p_1 + p_2 - p_3 - p_4\big)\,
\big|\psi_n(Q_1)\big|^2 
\big|\psi_n(Q_2)\big|^2
\big|\mathcal{M}\big|^2 
f_1^{\text{eq}} f_2^{\text{eq}}, \\[6pt]
&= 
\frac{1}{n_1^{\text{eq}} n_2^{\text{eq}}}
\int 
\frac{g_{B_n}\, d^3p_1}{(2\pi)^3 2E_1}
\frac{g_{B_n}\, d^3p_2}{(2\pi)^3 2E_2}
(\sigma v)_F \,
f_1^{\text{eq}} f_2^{\text{eq}},
\end{align*}

\begin{equation}
\langle \sigma v \rangle_{F} = 
\left[
\frac{K_1(M_{B_n}/T)}{K_2(M_{B_n}/T)}
\right]^{2}
(\sigma v)_F ~ \approx (\sigma v)_{F}.
\label{thcs_F}
\end{equation}
Another related process is,
\begin{equation*}
  \chi^*(p_1) + \chi^*(p_2) + B_n(p_3) \rightarrow B_n(p_4) + \Phi(p_5) \, .
\end{equation*}
The thermally averaged cross section for this process is evaluated as follows,

\begin{align*}
\langle \sigma v^2 \rangle_{G} 
&= 
\frac{1}{n_1^{\text{eq}} n_2^{\text{eq}} n_3^{\text{eq}}}
\left( \frac{2}{\mu} \right)^\frac{3}{2}
\int 
\frac{g_\chi\, d^3p_1}{(2\pi)^3 2E_1}
\frac{g_\chi\, d^3p_2}{(2\pi)^3 2E_2}
\frac{g_{B_n}\, d^3p_3}{(2\pi)^3 2E_3}
\frac{g_{B_n}\, d^3p_4}{(2\pi)^3 2E_4}
\frac{g_{\Phi}\, d^3p_5}{(2\pi)^3 2E_5} \\[0.5em]
&\quad \times 
\frac{d^3Q_1}{(2\pi)^3}
\frac{d^3Q_2}{(2\pi)^3}
\frac{d^3Q_3}{(2\pi)^3}
(2\pi)^4 \delta^4(p_1 + p_2 + p_3 - p_4-p_5)\,
|\psi_n(Q_1)|^2\, |\psi_s(Q_2)|^2\,|\psi_n(Q_3)|^2\,
|\mathcal{M}|^2 \\[0.5em]
&\quad \times
f_1^{\text{eq}} f_2^{\text{eq}} f_3^{\text{eq}}, \\[1em]
&= 
\frac{1}{n_1^{\text{eq}} n_2^{\text{eq}} n_3^{\text{eq}}}
\int 
\frac{g_\chi\, d^3p_1}{(2\pi)^3 2E_1}
\frac{g_\chi\, d^3p_2}{(2\pi)^3 2E_2}
\frac{g_{B_n}\, d^3p_3}{(2\pi)^3 2E_3}
(\sigma v^2)_G\,
f_1^{\text{eq}} f_2^{\text{eq}} f_3^{\text{eq}},
\end{align*}

\begin{equation}
  \langle \sigma v^2 \rangle_{G} = \left( \frac{K_1(m_\chi/T)}{K_2(m_\chi/T)} \right)^{\!2} \frac{K_1(M_{B_n}/T)}{K_2(M_{B_n}/T)}
  (\sigma v^2)_{G} ~ \approx (\sigma v^2)_{G}.
\end{equation}
Considering the Bose enhancement, the above equation becomes,
\begin{equation} 
    \langle \sigma v^2 (1 + f_A(\omega))\rangle_{G} 
    = (\sigma v^2)_{G}\, \bar{S}_{\rm BSF}(z),
    \label{eq:thcs_G_bose}
\end{equation}

The presence of the four point interaction term \(|\chi|^4\) also allows bound states to undergo number changing processes such as,
\begin{equation*}
  B_n(p_1) B_n(p_2) \rightarrow B_n(p_3) \Phi (p_4),
\end{equation*}  
\begin{align*}
\langle \sigma v \rangle_{H} 
&= 
\frac{1}{n_1^{\text{eq}} n_2^{\text{eq}} }
\left( \frac{2}{\mu} \right)^{\!3/2}
\int 
\frac{g_{B_n}\, d^3p_1}{(2\pi)^3 2E_1}
\frac{g_{B_n}\, d^3p_2}{(2\pi)^3 2E_2}
\frac{g_{B_n}\, d^3p_3}{(2\pi)^3 2E_3}
\frac{g_{\Phi}\, d^3p_4}{(2\pi)^3 2E_4} \\[0.5em]
&\quad \times 
\frac{d^3Q_1}{(2\pi)^3}
\frac{d^3Q_2}{(2\pi)^3}
\frac{d^3Q_3}{(2\pi)^3}
(2\pi)^4 \delta^4(p_1 + p_2 - p_3 - p_4)\,
|\psi_n(Q_1)|^2\, |\psi_s(Q_2)|^2\, |\psi_n(Q_3)|^2\,
|\mathcal{M}|^2 \\[0.5em]
&\quad \times
f_1^{\text{eq}} f_2^{\text{eq}},  \\[1em]
&= 
\frac{1}{n_1^{\text{eq}} n_2^{\text{eq}}}
\int 
\frac{g_{B_n}\, d^3p_1}{(2\pi)^3 2E_1}
\frac{g_{B_n}\, d^3p_2}{(2\pi)^3 2E_2}
(\sigma v)_H\,
f_1^{\text{eq}} f_2^{\text{eq}},
\end{align*}

\begin{equation}
  \langle \sigma v \rangle_{H} = \left( \frac{K_1(M_{B_n}/T)}{K_2(M_{B_n}/T)} \right)^{\!2} 
  (\sigma v)_{H}~ \approx (\sigma v)_{H}.
\end{equation}

Including the Bose enhancement, the above equation becomes,
\begin{equation}
    \langle \sigma v (1 + f_A(\omega))\rangle_{H} 
    = (\sigma v)_{H}\, \bar{S}_{\rm BSF}(z),
    \label{eq:thcs_H_bose}
\end{equation}

\subsection{Bound state dynamics cross sections}
The thermal average cross section for bound state formation process  $\chi(p_1) \chi(p_2) \rightarrow B_n(p_3) \Phi (p_4)$ will be,
\begin{align*}
    \langle \sigma v \rangle_{A} &= \frac{1}{n_1^{eq}n_2^{eq}} \sqrt{\frac{2}{\mu}} \int \frac{d^3Q_1}{(2\pi)^3}
\frac{d^3Q_2}{(2\pi)^3} \frac{g_\chi d^3p_1}{(2\pi)^3 2 E_1} \frac{g_\chi d^3p_2}{(2\pi)^3 2 E_2} \frac{g_B d^3p_3}{(2\pi)^3 2 E_3} \frac{g_{\Phi} d^3p_4}{(2\pi)^3 2 E_4} \\
    &\quad |\psi_n(Q_1)|^2\, |\psi_s(Q_2)|^2 (2\pi)^4 \delta^4(p_1+p_2-p_3-p_4) |M|^2 f_1^{eq} f_2^{eq}
\end{align*}
 Wwich would simplify as,
 \begin{equation}
  \langle \sigma v \rangle_{A}=\frac{1}{n_1^{eq}n_2^{eq}} \int \frac{g_\chi d^3p_1}{(2\pi)^3 2E_1 } \frac{g_\chi d^3p_2}{(2\pi)^3 2 E_2} (\sigma v)_{BSF} f_1^{eq} f_2^{eq},
  \label{thcs_bsf} 
\end{equation}

\begin{equation}
  \langle \sigma v \rangle_{A} =
  \left( \frac{K_1(m_\chi/T)}{K_2(m_\chi/T)} \right)^{\!2}
  (\sigma v)_{BSF} ~ \approx (\sigma v)_{BSF}.
\end{equation}

Including the Bose enhancement, the above equation becomes,
\begin{equation}
    \langle \sigma v(1 + f_A(\omega))\rangle_{A} 
    = (\sigma v)_{A}\, \bar{S}_{\rm BSF}(z),
    \label{eq:thcs_A_bose}
\end{equation}

The thermally averaged decay of bound state 
$B_n(p_1) \rightarrow \Phi(p_2) + \Phi(p_3)$ is given by,
\begin{equation}
\langle \Gamma \rangle_I
= \frac{1}{n_{B_n}^{\rm eq}}
\int \frac{g_{B_n}\, d^3 p_1}{(2\pi)^3 2E_1}\,
\Gamma_I\, f_1^{\rm eq} \, ,
\label{thcs_bsf}
\end{equation}
which can be evaluated analytically as
\begin{equation}
\langle \Gamma \rangle_I
= \left( \frac{K_1(m_{B_n}/T)}{K_2(m_{B_n}/T)} \right)\Gamma_I \, ~ \approx \Gamma_{I}.
\end{equation}

The thermally averaged de exciatation  rate of excited bound states 
$B_{n'}(p_1) \rightarrow B_n(p_2) + \Phi(p_3)$ is given by,
\begin{equation}
\langle \Gamma \rangle_J
= \frac{1}{n_{B_{n'}}^{\rm eq}}
\int \frac{g_{B_{n'}}\, d^3 p_1}{(2\pi)^3 2E_1}\,
\Gamma_I\, f_1^{\rm eq} \, ,
\label{thcs_bsf}
\end{equation}
which can be evaluated analytically as,
\begin{equation}
\langle \Gamma \rangle_J
= \left( \frac{K_1(m_{B_{n'}}/T)}{K_2(m_{B_{n'}}/T)} \right)\Gamma_J \, ~ \approx \Gamma_{J}.
\end{equation}

The thermal averaged cross section of ionization of SIMPonium would be given by the Milne relation,
\begin{equation*}
\langle \sigma v \rangle_{B} = \langle \sigma v \rangle_{A}  \frac{(n_\chi^{eq})^2}{n_B^{eq}},
\end{equation*}
\begin{equation}
\langle \sigma v \rangle_{B} = \langle \sigma v \rangle_{A}  \frac{m_\chi T}{4\pi} e^{-|E_n|/T},
\end{equation}
where $E_n$ is the binding energy of SIMPonium.

In the previous calculations, we employed the formal method to obtain the thermally averaged cross section. Here, we illustrate a simpler approach by considering the  process: $ \chi(p_1) \chi^*(p_2) \rightarrow A_\mu A_\mu $, for which the thermal average cross section 
is given by,
\begin{equation}
  \langle \sigma v \rangle_{K} = \frac{x^{3/2}}{2 \sqrt{\pi}}\int_{0}^{\infty} dv v^2  e^{\frac{-xv^2}{4}} (\sigma v)_{K}. 
\end{equation}
where,
\begin{equation*}
(\sigma v)_K = \frac{\pi^2 V_0^2}{ \mu^2} 
\frac{2\pi V_0}{v_{\mathrm{rel}}\left(1 - e^{-\frac{2\pi V_0}{v_{\mathrm{rel}}}}\right)}
\end{equation*}
  
The thermal average cross section for  annihilation of free $\chi$ into a pair of SM particles has a standard result given by,
\begin{equation}
   \langle \sigma v \rangle_{Ann} =\frac{x}{16 T m_\chi^4 K_2^2(x)}\int_{0}^{\infty} ds (\sigma v)_{Ann}K_1\!\bigg(\frac{\sqrt{s}}{T}\bigg)s\sqrt{s-4m_\chi^2}.
\end{equation}

\twocolumngrid
\bibliography{ref2}

\end{document}